\documentclass[aps,prd,groupedaddress,preprint,eqsecnum,nofootinbib,showpacs]{revtex4}
\usepackage{graphicx,epsf}
\usepackage[usenames]{color}
\newif\ifdraft
\drafttrue
\newif\ifpreprint
\preprinttrue

\def\sect#1{section~{\ref{#1}}}
\def\fig#1{fig.~{\ref{#1}}}

\def\spa#1.#2{\left\langle#1\,#2\right\rangle}
\def\spb#1.#2{\left[#1\,#2\right]}
\def\spash#1.#2{\spa{\smash{#1}}.{\smash{#2}}}
\def\spbsh#1.#2{\spb{\smash{#1}}.{\smash{#2}}}
\def\sand#1.#2.#3{%
\left\langle\smash{#1}{\vphantom1}^{-}\right|{#2}%
\left|\smash{#3}{\vphantom1}^{-}\right\rangle}
\def\sandpp#1.#2.#3{%
\left\langle\smash{#1}{\vphantom1}^{+}\right|{#2}%
\left|\smash{#3}{\vphantom1}^{+}\right\rangle}
\def\sandpm#1.#2.#3{%
\left\langle\smash{#1}{\vphantom1}^{+}\right|{#2}%
\left|\smash{#3}{\vphantom1}^{-}\right\rangle}
\def\sandmp#1.#2.#3{%
\left\langle\smash{#1}{\vphantom1}^{-}\right|{#2}%
\left|\smash{#3}{\vphantom1}^{+}\right\rangle}

\def\twoloop{{2 \mbox{-} \rm loop}}

\def\tree{{\rm tree}}
\def\pol{\varepsilon}
\def\Tr{\, {\rm Tr}}

\def\eps{\epsilon}

\def\nn{\nonumber}

\def\eqn#1{eq.~(\ref{#1})}

\def\eqns#1#2{eqs.~(\ref{#1}) and~(\ref{#2})}

\def\NeqFour{{{\cal N}=4}}

\def\NeqEight{{{\cal N}=8}}

\def\f{\widetilde f}
\def\be{\begin{equation}}
\def\ee{\end{equation}}
\def\bea{\begin{eqnarray}}
\def\eea{\end{eqnarray}}
\def\ba{\begin{eqnarray}}
\def\ea{\end{eqnarray}}

\def\Perm{{\cal P}}
\def\M{{\cal M}}

\def\P{{\rm P}}
\def\NP{{\rm NP}}
\def\mud{\lambda}
\def\bowtie{{\rm bow\mbox{-}tie}}

\def\P{{\rm P}}
\def\NP{{\rm NP}}
\def\n{\widetilde n}

\def\tree{{\rm tree}}

\newbox\charbox
\newbox\slabox
\def\s#1{{      % Feynman slash
        \setbox\charbox=\hbox{$#1$}
        \setbox\slabox=\hbox{$/$}
        \dimen\charbox=\ht\slabox
        \advance\dimen\charbox by -\dp\slabox
        \advance\dimen\charbox by -\ht\charbox
        \advance\dimen\charbox by \dp\charbox
        \divide\dimen\charbox by 2
        \raise-\dimen\charbox\hbox to \wd\charbox{\hss/\hss}
        \llap{$#1$} }}

\DeclareMathAlphabet{\mathpzc}{OT1}{pzc}{m}{it}

\begin{document}

\ifpreprint
\hfill UCLA/07/TEP/15
\fi

\title{New Relations for Gauge-Theory Amplitudes}

\author{Z.~Bern, J.~J.~M.~Carrasco and H.~Johansson}

\affiliation{
Department of Physics and Astronomy, UCLA, Los Angeles, CA
90095-1547, USA 
}

%\date{May 25, 2008}

\begin{abstract}
We present an identity satisfied by the kinematic factors of diagrams
describing the tree amplitudes of massless gauge theories.  This
identity is a kinematic analog of the Jacobi identity for color
factors.  Using this we find new relations between color-ordered
partial amplitudes.  We discuss applications to multiloop
calculations via the unitarity method.  In particular, we illustrate
the relations between different contributions to a two-loop four-point
QCD amplitude.  We also use this identity to reorganize gravity tree
amplitudes diagram by diagram, offering new insight into the structure
of the KLT relations between gauge and gravity tree amplitudes. This
can be used to obtain novel relations similar to the KLT ones.  We
expect this to be helpful in higher-loop studies of the ultraviolet
properties of gravity theories.
\end{abstract}

\pacs{11.15.Bt, 11.25.Db, 11.25.Tq, 11.55.Bq, 12.38.Bx \hspace{1cm}}

\maketitle

%%%%%%%%%%%%%%%%%%%%%%%%%%%%%%%%
\section{Introduction}
\label{IntroSection}

Gauge and gravity scattering amplitudes have a far simpler and richer
structure than apparent from Feynman rules or from their respective
Lagrangians.  Striking tree-level examples include the Parke-Taylor
maximally helicity violating (MHV) amplitudes in
QCD~\cite{ParkeTaylor}, the delta-function support of amplitudes on
polynomial curves in twistor space~\cite{WittenTopologicalString,
RoibanString}, and the Kawai-Lewellen-Tye (KLT)
relations~\cite{KLT,KLTGeneral,Freedman1} between gravity and gauge-theory 
tree amplitudes.  Besides their intrinsic theoretical value,
these structures have led to a number of useful computational
advances, as described in various reviews~\cite{TreeReview,
OneLoopReview, GravityReview, CSReview}.

In particular, tree-level color-ordered partial amplitudes
satisfy simplifying relations dictated by the color algebra.  
Adjoint representation amplitudes must vanish whenever an
external gluon is replaced by a color neutral photon, giving a
``photon decoupling identity''~\cite{TreeReview} (also referred to as
the subcyclic property).  More generally, the
Kleiss-Kuijf relations~\cite{KleissKuijf,LanceColor} reduce the number
of independent $n$-point tree partial amplitudes from $(n-1)!$ to
$(n-2)!$ partial amplitudes.

In this paper we propose a kinematic identity that further constrains
the $n$-point color-ordered partial amplitudes at tree-level.  This is
based on the observation that gauge-theory amplitudes can be
rearranged into a form where the kinematic factors of
diagrams describing the amplitudes satisfy an identity analogous to
the Jacobi identity obeyed by the color factors associated with the
same diagrams. At four points this kinematic identity has been
used previously to explain certain zeroes in cross
sections~\cite{Halzen}. 
By solving the generated set of equations at higher-points, we obtain
new nontrivial relations amongst color-ordered tree amplitudes,
reducing the number of independent partial amplitudes to $(n-3)!$.

The existence of such an identity is unobvious from the Feynman
diagrams contributing to higher-point tree amplitudes.  Indeed,
color-ordered Feynman diagrams, in a generic choice of gauge, will
{\it not} satisfy this new identity in isolation beyond four points.
Rather the kinematic factors satisfying the identity only appear after
rearranging terms between contributing Feynman diagrams into
convenient representations of the amplitudes.  In this paper we do not
present a complete proof that this rearrangement is always possible.
However, because of the many explicit tree-level checks that we have
performed, and because of its close connection to the color Jacobi
identity, we expect this new $n$-point identity to hold for 
tree-level color-ordered Yang-Mills amplitudes.  In this paper we
will focus on gluonic amplitudes.

Although the amplitude relations derived from the new kinematic identity
may be helpful in tree and one-loop calculations, powerful
computational methods are already available, or are under
development~\cite{TreeReview,OneLoopReview,CSReview,LesHouches}.  In
the past decade important progress has also been made in computing
higher-loop scattering amplitudes both for phenomenological and
theoretical purposes.  For example, on the phenomenological side, even
fully differential cross-sections for processes as complicated as $e^+
e^- \rightarrow 3$ jets at next-to-next-to-leading order are now
computable~\cite{eeTwoloop}.  Much of this progress relies on various
improvements in loop integration techniques~\cite{MultiloopIntegrals}.
On the theoretical side, multiloop calculations of scattering
amplitudes have become important as a means of studying fundamental
issues in gauge and gravity
theories~\cite{ABDK,BCDKS,Finite,GravityThree}.  Various useful
relations aiding computations in higher-loop maximally supersymmetric
theories have been discussed in refs.~\cite{BRY, BDDPR, Sokatchev,
BCDKS, FiveLoop, Freddy}.  To go beyond this, a greatly improved
understanding of the structure of multiloop scattering amplitudes
will likely be important.

The unitarity method~\cite{UnitarityMethod}, gives us a means of
transferring properties of amplitudes from tree level to loop level.
Since this approach constructs loop diagrams out of tree-level
amplitudes, we can apply the relations following from the kinematic identity
to help simplify multiloop calculations.  Specifically, we will show
that it induces nontrivial relations between planar and nonplanar
loop-level contributions.  Its application to the construction of the
four-loop four-point amplitude of $\NeqFour$ super-Yang-Mills
theory in terms of loop integrals---including nonplanar
contributions---will be given elsewhere~\cite{FourLoopNonplanar}.  The
planar contributions at four and five loops have already been given in
ref.~\cite{BCDKS,FiveLoop}.

Besides applications to gauge theories, the identity also suggests
a natural reorganization of gravity tree amplitudes so that the
numerator of each kinematic pole in the amplitude is given by a
product of two gauge-theory kinematic numerators.  As we will show,
this is closely connected to the KLT relations between color-ordered
gauge-theory and gravity amplitudes.  The new representations for
gravity tree amplitudes can be used in loop calculations via
generalized unitarity~\cite{BDDPR, GeneralizedUnitarity, Finite,
GravityThree}.  There may also be a connection to other recently
uncovered relations between maximally helicity violating gravity and
gauge-theory amplitudes~\cite{Freedman1}, but this requires further
study.

This paper is organized as follows.  In \sect{ReviewSection} we review
some pertinent properties needed later in the paper, as well as
establish notation.  In \sect{FourPoint}, 
we derive the identity at four points, guiding our higher-point construction.
Then in \sect{NPointTwist} we motivate the higher-point
generalization of the identity, discussing the five-point case in
some detail.  In \sect{HigherLoopSection} we work out multiloop QCD
examples of its application.  In \sect{GravitySection} we present
implications for gravity amplitudes, showing how the identity
clarifies the KLT relations and can be used to derive new
representations for gravity tree amplitudes in terms of gauge-theory
ones.

\section{Review}
\label{ReviewSection}

In this section we set up the terminology, notation and review a
number of pertinent results directly relevant for our subsequent
discussion.

\subsection{Gauge-theory color structure}
\label{ColorSubsection}

At tree level, with particles all in the adjoint
representation of $SU(N_c)$, the full tree amplitude can be decomposed as
\begin{equation}
{\cal A}^\tree_n (1,2,3, \ldots, n)=g^{n-2} 
\sum_{\Perm (2,3,\ldots, n)} {\rm Tr}[T^{a_1}T^{a_2} T^{a_3}\cdots T^{a_n}] 
\, A^\tree_n (1,2,3, \ldots, n),
\label{TreeDecomposition}
\end{equation}
where $A_n^\tree$ is a tree-level color-ordered $n$-leg partial
amplitude. The $T^a$'s are color-group generators, encoding the color
of each external leg $1,2,3 \ldots n$. The sum is over all noncyclic
permutations of legs, which is equivalent to all permutations keeping
leg $1$ fixed.  Helicities and polarizations are suppressed.  At
higher loops, detailed color decompositions of gauge-theory amplitudes
have not been given though general properties are clear.  For example,
at $L$ loops the corresponding decomposition involves up to $L +1$
color traces per term.  Discussions of such color decompositions at
tree level and at one loop may be found in
refs.~\cite{TreeReview,OneloopReview}.

%%%%%%%%% FIGURE %%%%%%%%%%%%%%%%%%
\begin{figure}[t]
\centerline{\epsfxsize 3.5 truein \epsfbox{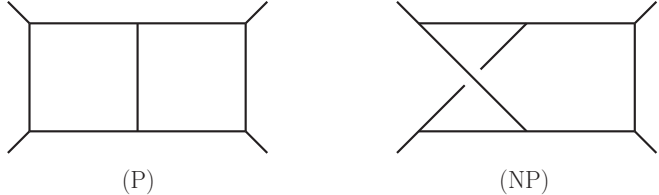}}
\caption[a]{\small The ``parent'' diagrams for the two-loop four-point
identical-helicity amplitudes of QCD and $\NeqFour$ super-Yang-Mills
theory. All other diagrams appearing in the amplitude are obtained by
collapsing propagators.  These parent diagrams also
determine the color factors appearing in \eqn{FColor}, by dressing
them with $\f^{abc}$s in a clockwise ordering.}
\label{TwoLoopParentsFigure}
\end{figure}
%%%%%%%%%%%%%%%%%%%%%%%%%%%%%%%%

Other color decompositions involve using the $f^{abc}$ group
structure constants.  At tree level and at one loop 
a decomposition of this type was given in
refs.~\cite{DelDucaColor,LanceColor}.  For tree level, this
decomposition is similar to the one in \eqn{TreeDecomposition} using,
instead, color matrices in the adjoint representation (the
$f^{abc}$'s), and summing over fewer partial amplitudes.

Color-ordered tree-level amplitudes satisfy a set of well-known relations. 
The simplest of these are the cyclic  and reflection properties,
\begin{equation}
A_n^\tree(1,2, \ldots, n) = A_n^\tree(2, \ldots, n,1) \,,  
\hskip 0.7 cm  A_n^\tree(1,2, \ldots, n) = (-1)^n A_n^\tree(n,\ldots,2,1) \,.
\end{equation}
Next there is the ``photon''-decoupling identity 
(or subcyclic property)~\cite{TreeReview, KleissKuijf}, 
\begin{equation}
\sum_{\sigma \in \rm cyclic} A_n^\tree(1,\sigma(2,3, \ldots, n) ) = 0\,,
\label{Decoupling}
\end{equation}
where the sum runs over all cyclic permutations of legs $2,3,4, \ldots
n$.  This identity follows by replacing $T^{a_1} \rightarrow 1$
in the full amplitude (\ref{TreeDecomposition}) corresponding to
replacing leg $1$ with a ``photon''.  The amplitude must then vanish
since photons cannot couple directly to adjoint representation
particles.

Other important relations are the Kleiss-Kuijf
relations~\cite{KleissKuijf}:
\begin{equation}
A_n^\tree(1,\{\alpha\}, n, \{\beta\})=(-1)^{n_\beta}\sum_{\{\sigma\}_i \in {\rm OP}(\{\alpha\}, \{\beta^T\})}  A_n^\tree(1,\{\sigma\} _i,n ) \,,
\label{KleissKuijf}
\end{equation}
where the sum is over the ``ordered permutations'' ${\rm
OP}(\{\alpha\}, \{\beta^T\})$, that is, all permutations of
$\{\alpha\} \bigcup \{\beta^T\}$ that maintains the order of the
individual elements belonging to each set within the joint set.  The
notation $\{\beta^T\}$ represents the set $\{\beta\}$ with the
ordering reversed, and $n_\beta$ is the number of $\beta$ elements.
These relations were conjectured in ref.~\cite{KleissKuijf} and proven in
ref.~\cite{LanceColor}.

Consider, as an example, a five-point tree amplitude.  For
$A_5^\tree(1,\{2,3\},5,\{4\})$ we have,
\begin{equation}
A_5^\tree(1,2,3,5,4) = -A_5^\tree(1,2,3,4,5) -A_5^\tree(1,2,4,3,5) 
- A_5^\tree(1,4,2,3,5)  \,.
\label{KleissKuijfFiveRelation}
\end{equation}
The other five-point relations are given by permuting legs $2,3,4$ and
using the cyclic and reflection properties.  This means that the six
amplitudes $A_5^\tree(1,\Perm\{2,3,4\},5)$---where $\Perm\{2,3,4\}$ signifies
all permutations over legs $\{2,3,4\}$---form a basis in which the
remaining five-point partial amplitudes can be expressed. More generally,
for multiplicity $n$, the Kleiss-Kuijf relations can be used to rewrite
any color-ordered partial amplitude in terms of only $(n-2)!$ 
basis partial amplitudes, where two legs are held fixed in the ordering.

For the two-loop four-point identical-helicity pure gluon and two-loop
MHV amplitudes in maximally supersymmetric Yang-Mills theory, a color
decomposition in terms of $f^{abc}$ has been given in terms of
``parent'' diagrams establishing both the color and kinematic
structure~\cite{BDDPR,TwoLoopAllPlus},
\begin{equation}
{\cal A}_4^\twoloop(1,2,3,4) = g^6  
\Bigl[ C^\P_{1234} \, A^\P_{1234} + C^\P_{3421} \, A^\P_{3421}
   + C^\NP_{12;34} \, A^\NP_{12;34} + C^\NP_{34;21} \, A^\NP_{34;21}
    +\ {\cal C}(234) \Bigr] \,,
\label{FColor}
\end{equation}
where ``$+\ {\cal C}(234)$'' signifies that one should add the two
cyclic permutations of $2,3,4$.  The $A^\P$ and $A^\NP$ are primitive
amplitudes stripped of color.  The values of the color coefficients
$C^\P$ and $C^\NP$ may be read off directly from the parent diagrams in
\fig{TwoLoopParentsFigure}.  For example, $C^\P_{1234}$ is the color
factor obtained from diagram (a) by dressing each vertex with an
$\f^{abc}$, where
\begin{equation}
\f^{abc} \equiv i\sqrt{2} f^{abc} = \Tr\bigl( [T^a,T^b] T^c \bigr),
\label{ftildedef}
\end{equation}
and dressing each internal line with a $\delta^{ab}$.  In
\sect{HigherLoopSection}, we will make use of this decomposition in
a two-loop example.  As discussed in ref.~\cite{LanceColor},
general representations in terms of parent diagrams can be
constructed by making repeated use of the Jacobi identity, but this
has not been studied in full generality.

%%%%%%%%% FIGURE %%%%%%%%%%%%%%%%%%
\begin{figure}[t]
\centerline{\epsfxsize 5 truein \epsfbox{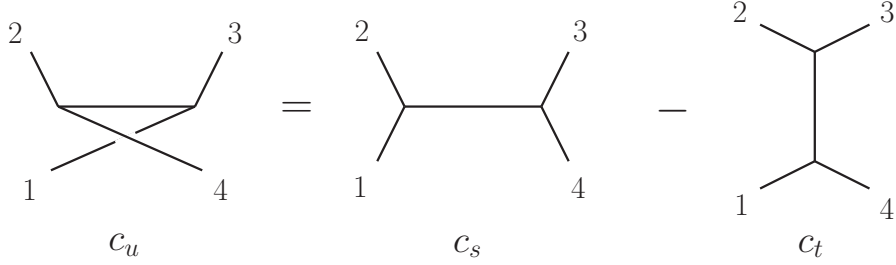}}
\caption[a]{\small The Jacobi identity relating the color factors of
the $u, s, t$ channel ``color diagrams''. The color factors are given by dressing
each vertex with an $\f^{abc}$ following a clockwise ordering.}
\label{ColorJacobiFigure}
\end{figure}
%%%%%%%%%%%%%%%%%%%%%%%%%%%%%%%%

A key property of the $\f^{abc}$'s are that they satisfy the Jacobi
identity illustrated in \fig{ColorJacobiFigure},
\begin{eqnarray}
&& c_u \equiv \f^{a_4 a_2 b} \f^{b a_3 a_1 } \,, \hskip 1 cm 
c_s \equiv \f^{a_1 a_2 b} \f^{b a_3 a_4} \,, \hskip 1 cm 
c_t \equiv \f^{a_2 a_3 b} \f^{b a_4 a_1} \,, \nn \\
&& \hskip 4.4 cm 
c_u = c_s - c_t\,.
\label{JacobiIdentity}
\end{eqnarray}
The main result of this paper is that the kinematic factors
corresponding to $n$-point tree diagrams can be rearranged to
satisfy an analogous identity giving nontrivial constraints on the
form of tree amplitudes.  This then has useful consequences at loop
level and for corresponding gravity amplitudes.

\subsection{Higher-loop Integral representation and the unitarity method}

Any multiloop amplitude can be expanded in a set of loop integrals
with rational coefficients,
\begin{equation}
A_n^{(L)} =\sum_{i} a_i I_n^{(L),i}\, ,
\end{equation}
where the $a$'s are the coefficients, $i$ runs over a list of
integrals $I_n^{(L),i}$, and $A_n^{(L)}$ is a generic $n$-point $L$-loop 
amplitude, not necessarily color decomposed.  Here, we neither demand
that the integrals form a linear independent basis under integral
reductions, nor that integrals vanishing under integration be removed
from the set.

In this paper we consider representations of $n$-point $L$-loop
integrals that can be written in a $D$-dimensional Feynman-like
manner, schematically,
\begin{equation}
I_n^{(L),i} =
\int \Bigl( \prod_{m=1}^L {d^D  l_m\over (2\pi)^D} \Bigr) \,  
{N_i \over \prod_j p_j^2},
\end{equation}
where the propagators, $1/p^2_j$, specify the structure of the graph.
The numerator factors $N_i$ in general depend on the loop momenta and
also on external kinematics and polarizations.  The number of powers
of loop momenta that can appear in the numerator factors depends on
the theory under consideration.  The $l_m$ are the $L$ independent
variables of loop momenta, usually picked from the set of propagator
momenta $p_j$.

%%%%%%%%% FIGURE %%%%%%%%%%%%%%%%%%
\begin{figure}[t]
\centerline{\epsfxsize 2.6 truein \epsfbox{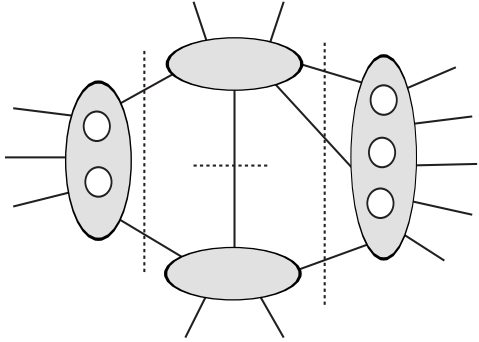}}
\caption[a]{\small The unitarity method constructs multiloop
amplitudes from lower-loop amplitudes.  The blobs represent
amplitudes, the white holes loops and the dotted lines indicate cuts
which replace propagators with on-shell delta functions. Generalized
cuts which decompose loop amplitude solely in terms of tree amplitudes
are particularly useful in carrying out multiloop calculations.}
\label{GeneralizedCutExample}
\end{figure}
%%%%%%%%%%%%%%%%%%%%%%%%%%%%%%%%

To carry over the newly uncovered tree-level relations to loop level
we use the unitarity method~\cite{UnitarityMethod,
GeneralizedUnitarity, OneloopReview, BRY, BDDPR, TwoLoopSplit} to
construct complete loop-level amplitudes, at the level of the
integrands, prior to carrying out any loop integration.  In this
method higher-loop-level integrands are
constructed by taking the product of lower-loop or tree amplitudes
and imposing on-shell conditions on intermediate
legs. As illustrated in \fig{GeneralizedCutExample}, the unitarity cuts
are given by a product of lower-loop amplitudes,
\begin{equation}
\sum_{\rm states} A_{(1)}A_{(2)} A_{(3)} \cdots  A_{(m)} \,,
\label{GeneralizedCut}
\end{equation}
where the sum runs over all particle types and physical states that
can propagate  on the internal cut lines.  A complete
integrand is then found by systematically constructing an ansatz that
has the correct cuts in all channels.  Overviews of the unitarity
method may be found in refs.~\cite{OneloopReview}.  A systematic
description of the merging procedure for constructing multiloop
amplitudes from the cuts may be found in ref.~\cite{TwoLoopSplit}.
Examples of explicit higher-loop calculations using the unitarity
method may be found in refs.~\cite{BRY,BDDPR}.

For the purposes of this paper, the generalized cuts gives us a means
of applying the new tree-level relations and identities directly at
higher-loop level, since it allows us to construct amplitudes entirely
from tree-level amplitudes.

A particularly useful set of cuts are the maximal
cuts~\cite{BCFGeneralized, FiveLoop, Freddy} where the maximum number
of propagators are put on-shell\footnote{We use the terminology of
ref.~\cite{FiveLoop}, not refs.~\cite{Freddy} where maximal cuts
include additional hidden singularities as well.}.  In this case, the
cut is given by a sum over products between three-point tree
amplitudes, isolating a particular parent diagram (modulo any contact
terms).  We build complete amplitudes by systematically releasing
on-shell conditions to identify all relevant contact terms.  This
procedure has been described in some detail in ref.~\cite{FiveLoop}.
Such maximal cuts may be exploited in $D$-dimension, as we do in
\sect{HigherLoopSection}.

\subsection{Gravity amplitudes}

At tree level, gravity amplitudes satisfy a remarkable relation with
gauge-theory amplitudes, first uncovered in string theory by Kawai,
Lewellen and Tye~\cite{KLT,KLTGeneral,Freedman1}.  These relations
also hold in field theory, as the low-energy limit of string theory.
In this limit, the KLT relations for four-, five- and
six-point amplitudes are,
\begin{eqnarray}
M_4^\tree (1,2,3,4) & =& - i s_{12} A_4^\tree(1,2,3,4) \,
   \widetilde{A}_4^\tree(1,2,4,3)\,, \label{KLT4}\\
M_5^\tree(1,2,3,4,5) 
&=& i s_{12} s_{34}  A_5^\tree(1,2,3,4,5) \,
                \widetilde{A}_5^\tree(2,1,4,3,5) \nn \\
&& \hskip 2 cm 
+ i s_{13}s_{24} A_5^\tree(1,3,2,4,5) \, 
           \widetilde{A}_5^\tree(3,1,4,2,5) \,,\label{KLT5}\\
M_6^\tree(1,2,3,4,5,6) 
& =& - i s_{12} s_{45}  A_6^\tree(1,2,3,4,5,6) 
    [ s_{35} \widetilde{A}_6^\tree(2,1,5,3,4,6)  \nn\\
&& \null \hskip 4 cm 
   + (s_{34} + s_{35}) \widetilde{A}_6^\tree(2,1,5,4,3,6) ] \nn \\
&& \hskip 2 cm \null
+\ \Perm(2,3,4) \,.
\label{KLT6}
\end{eqnarray}
Here the $M_n$'s are amplitudes in a gravity theory stripped of
couplings, the $A_n$'s and $\widetilde{A}_n$'s are the color-ordered
amplitudes in two, possibly different, gauge theories (the gravity
states are direct products of gauge-theory states for each external
leg)~\cite{TreeColor,TreeReview}, $s_{ij}\equiv s_{i,j}= (k_i+k_j)^2$
with $k_i$ being the outgoing momentum of leg $i$, and $\Perm(2,3,4)$
signifies a sum over all permutations of the labels 2, 3 and 4.
An $n$-point generalization of the KLT relations
is~\cite{MHVGravityOneLoop},
\begin{eqnarray}
M^\tree_n(1,2,\ldots,n)&=&i(-1)^{n+1}
\Bigl[A^\tree_n(1,2,\ldots,n)\displaystyle 
\sum_{\rm perms}f(i_1,\ldots,i_j)\overline{f}(l_1,\ldots,l_{j'}) \nn\\
&&\hskip 3 cm \null\times 
\widetilde{A}^\tree_n(i_1,\ldots,i_j,1,n-1,l_1,\ldots,l_{j'},n)\Bigr]\nn\\
&& \null \hskip 1 cm 
 +\Perm(2,\ldots,n-2) \,, 
\label{KLTnpoint}
\end{eqnarray}
where the sum is over all permutations $\{i_1,\ldots,i_j\} \in
\Perm\{2,\ldots,\lfloor n/2\rfloor\} $ and $\{l_1,\ldots,l_{j'}\}\in
\Perm\{\lfloor n/2\rfloor+1,\ldots,n-2\} $ with $j=\lfloor n/2
\rfloor-1$ and $j'= \lfloor n/2 \rfloor-2$, which gives a total of
$(\lfloor n/2 \rfloor-1)! (\lfloor n/2 \rfloor-2)!$ terms inside the
square brackets. The notation ``$+\,\Perm(2,\ldots,n-2)$'' signifies
a sum over the preceding expression for all permutations of legs
$2,\ldots,n-2$. The functions $f$ and $\overline{f}$ are given by,
\begin{eqnarray}
f(i_1,\ldots,i_j)&=&s_{1,i_j}\prod_{m=1}^{j-1}
\left(s_{1,i_m}+\sum_{k=m+1}^{j} g(i_m,i_k) \right),\nn\\
\overline{f}(l_1,\ldots,l_{j'})&=&s_{l_1,n-1}\prod_{m=2}^{j'}
\left(s_{l_m,n-1}+\sum_{k=1}^{m-1} g(l_k,l_m) \right) \,,
\end{eqnarray}
and the function $g$ is,
\begin{equation}
g(i,j)=\left\{ 
\begin{array}{ll}
        s_{i,j}  & \mbox{if $i> j$}\\
        0& \mbox{else}
\end{array} \right\} \,.
\end{equation}
The full gravity amplitudes are obtained by multiplying 
with gravity coupling constants,
\begin{equation}
\M^\tree_n= \Bigl( {\kappa \over 2} \Bigr)^{n-2} M^\tree_n \,.
\end{equation}
%

%%%%%%%%%%%%%%%%%%%%%%%%%%

\section{An identity at four points}
\label{FourPoint}

In this section we discuss an identity that kinematic numerator
factors of color-ordered four-point gauge-theory amplitudes
satisfy. Specifically, we show that four-point gauge-theory amplitudes
can be decomposed in terms of numerators $n_s,n_t,n_u$ of kinematic
poles of the Mandelstam variables, $s$, $t$, and $u$.  As we shall see
these numerator factors satisfy an identity analogous to the Jacobi
identity (\ref{JacobiIdentity}) for adjoint representation color
factors.  For four-point tree amplitudes this identity may appear to
be a curiosity, but as we will see below, the consequences at higher
points and loops will be rather nontrivial.  Interestingly, the
identity had been noted almost three decades ago at the four-point
level~\cite{Halzen}.

To derive the identity we will utilize general properties of adjoint
representation gluonic amplitudes. As explained in
\sect{ColorSubsection}, color-ordered tree-level amplitudes satisfy a
photon-decoupling identity~\cite{TreeReview}. At four points we have,
\begin{equation}
A_4^\tree(1,2,3,4) + A_4^\tree(1,3,4,2) + A_4^\tree(1,4,2,3)=0 \,.
\label{DecouplingFour}
\end{equation}
To exploit this equation, we note that tree amplitudes in general are
rational functions of polarization vectors, spinors, momenta and
Mandelstam invariants, $s=(k_1 + k_2)^2$, $t=(k_1+k_4)^2$ and $u =
(k_1+k_3)^2$, where the $k_i$ are outgoing massless momenta,
corresponding to each external leg $i$. Because the decoupling
identity (\ref{DecouplingFour}) does not rely on the specific
polarizations or space-time dimension, the cancellation is entirely
due to the amplitude's dependence on the Mandelstam variables. In
particular, it cannot rely on four-dimensional spinor identities.  We
recognize that the only nontrivial way this can happen is for the sum
in \eqn{DecouplingFour} be equivalent to the vanishing of $s+t+u$ as
follows:
\begin{equation}
A_4^\tree(1,2,3,4) + A_4^\tree(1,3,4,2) + A_4^\tree(1,4,2,3) =
        {(s+t+u) \chi} = 0 \,,
\label{DecouplingImplication}
\end{equation}
where $\chi$ is a shared factor that depends on the polarizations
and momenta. From \eqn{DecouplingImplication}, the
partial amplitudes should be proportional to each other.  Furthermore,
since $A^\tree_4(1,2,3,4)$ treats any factors of $s$ and $t$ the same,
its contribution to \eqn{DecouplingImplication} should be proportional
to $u=-(s+t)$.  We therefore make the following identification,
\begin{equation}
A^\tree_4(1,2,3,4)=u\chi \,.
\end{equation}
Similar considerations give,
\begin{equation}
 A^\tree_4(1,3,4,2) = t \chi, \hskip 2 cm  A_4^\tree(1,4,2,3)=s\chi \,,
\end{equation}
consistent with \eqn{DecouplingImplication}.
After eliminating $\chi$ we obtain the following relations between
four point amplitudes,
\begin{eqnarray}
&& t\, A_4^\tree(1,2,3,4)=u\, A_4^\tree(1,3,4,2) \,,  
\hskip 1cm s\, A_4^\tree(1,2,3,4)=u\, A_4^\tree(1,4,2,3)\,, \nn \\ &&
t\, A_4^\tree(1,4,2,3)=s \, A_4^\tree(1,3,4,2)\,.
 \label{FourPointRelation}
\end{eqnarray}

We note that the well-known four-point tree-level helicity
amplitudes in $D=4$ explicitly satisfy this.  For example, pure gluon
amplitudes in the two color orders,
\begin{eqnarray}
&& A^\tree_4(1^-,2^-,3^+,4^+)=i\frac{{\spa1.2 }^4} 
{{\spa1.2 }{\spa2.3 }{\spa3.4 }{\spa4.1}}=
-i\frac{{\spa1.2 }^2{\spb 3.4 }^2} {st}  \,, \nn \\
&& A^\tree_4(1^-,4^+,2^-,3^+) =i\frac{{\spa1.2 }^4}
 {{\spa1.4 }{\spa4.2 }{\spa2.3 }{\spa3.1}}
=-i\frac{{\spa1.2 }^2{\spb 3.4 }^2} {tu}\, ,
\end{eqnarray}
satisfy the relations (\ref{FourPointRelation}). The $\spa{a}.{b}$ 
and $\spb{a}.{b}$ are spinor inner products of Weyl spinors, 
using notation of refs.~\cite{TreeReview}.

To obtain the kinematic analog of the Jacobi identity we exploit the
fact that the color-ordered tree amplitudes can be expanded in 
a convenient representation in terms of the poles that appear,
\begin{eqnarray}
A^\tree_4(1,2,3,4) &\equiv& {n_s \over s} + {n_t \over t} \,,\nn \\
A^\tree_4(1,3,4,2) &\equiv& -{n_u \over u} - {n_s \over s} \,,\nn \\
A^\tree_4(1,4,2,3) &\equiv& -{n_t \over t} + {n_u \over u} \,.
\label{PoleForm}
\end{eqnarray}
Practically, this can be done in terms of Feynman diagrams (by
absorbing the quartic contact terms into the cubic diagrams).
One may also think of the numerators as unknown until solved
for, and thus \eqn{PoleForm} defines the $n_i$'s. The sign flipping is
due to the antisymmetry of color-ordered Feynman rules.
(The overall signs of the numerators depend on our choice
of conventions.  Of course, trivial redefinitions of the type
$n_i\rightarrow-n_i$, can be used to modify the signs used to define
the $n_i$, but this cannot change the actual value of the diagram or
residues that the numerators correspond to.)
 
Comparing \eqn{FourPointRelation} to \eqn{PoleForm} gives us the
desired kinematic numerator identity,
\begin{equation}
n_u = n_s - n_t \,.
\label{TwistIdentity2}
\end{equation}
With the chosen sign conventions in \eqn{PoleForm}, this is 
exactly of the same form as the Jacobi identity for the color factors
given in \eqn{JacobiIdentity}.  

Using the definition of the fully dressed amplitude
\eqn{TreeDecomposition} and the partial amplitude form \eqn{PoleForm},
after converting to the $\f^{abc}$'s, 
we obtain a color-dressed representation,
\begin{equation}
{\cal A}^\tree_4 =
g^2 \biggl( \frac{n_s c_s}{s} + \frac{n_t c_t}{t} + 
              \frac{n_u c_u}{u} \biggr) \,,
\label{FullTree}
\end{equation}
where the color factors are given in (\ref{JacobiIdentity}). With our
sign conventions the signs in this color-dressed diagram
representation are all positive. Note that the form of \eqn{FullTree}
is very similar to the usual expansion in terms of Feynman diagrams,
except  that we decomposed the four-point contact terms according to
their color factors. In absorbing the contact terms into 
\eqn{FullTree} we must ensure that no cross terms, such as $n_sc_t/s$, appear.

We may take the kinematic numerators to be local in the external
polarizations and momenta. A natural question is whether these objects
are unique gauge-invariant quantities.  One check would be to consider
the limit $s \rightarrow 0$ of the $A_4^\tree(1,2,3,4)$ amplitude,
where the four-point amplitude factorizes into two three-point
amplitudes, each of which are manifestly gauge invariant.  From this
we would conclude that $n_s$ is gauge invariant in the $s\rightarrow
0$ factorization limit.  (To make $n_s$ nonvanishing in the limit we
may use complex momenta~\cite{ComplexMomenta,
WittenTopologicalString}.)  However, we should think of $n_s/s$ as
essentially an $s$-channel Feynman diagram, and of course individual
Feynman diagrams are not gauge invariant, so we would properly
conclude that $n_s$ is gauge dependent for nonzero values of $s$ and, more
generally, dependent on the choice of field variables.  Under the
assumption that $n_s$ is local, this freedom corresponds to all
possible terms that can be added to $n_s$ which cancel the $1/s$
pole. We will call this a ``gauge freedom'', because the gauge
transformations are a familiar concept corresponding to a freedom of
moving terms between diagrams.%
\footnote{We use the words ``gauge freedom'' loosely here; the freedom
does not necessarily mean that a gauge 
transformation exists that causes a particular rearrangement of terms.}
We may parametrize this freedom as,
\begin{equation}
n'_s=n_s + \alpha(k_i,\pol_i) s \,,
\label{SgaugeInvariance}
\end{equation}
where $\alpha(k_i,\pol_i)$ is local.  This corresponds to a contact
term ambiguity, and it does not change the residue of the
$s$-pole. However, to keep the value of the tree amplitudes in
\eqn{PoleForm} unchanged we must simultaneously change $n_t$ and
$n_u$,
\begin{equation}
n_t'=n_t-\alpha(k_i,\pol_i)t, \hskip 1 cm n_u'=n_u-\alpha(k_i,\pol_i)u\,.
\label{TUgaugeInvariance}
\end{equation}
Notice that this is exactly what is needed to make the sum of the
shifted numerators vanish,
\begin{equation}
-n_s'+n_t'+n_u'=-n_s+n_t+n_u-\alpha(k_i,\pol_i)(s+t+u)=0 \,.
\label{invariance}
\end{equation}
Therefore the transformation in \eqn{SgaugeInvariance} and
\eqn{TUgaugeInvariance} has the effect of moving contact terms between
the $s$, $t$ and $u$ channel diagrams without altering the numerator
identity~(\ref{TwistIdentity2}).  Although the numerators depend
on the gauge choices, the identity (\ref{TwistIdentity2}) remains true
for all gauges.  As we shall see in \sect{NPointTwist}, this property
is special to four points. At higher points, only specific choices of
numerators will satisfy the analogous identities and the form obtained
from generic gauge Feynman diagrams will not.

In general we can choose the numerators to be local.  It should be
noted, however, that even if we allow $\alpha(k_i,\pol_i)$ to be
nonlocal, the kinematic identity \eqn{TwistIdentity2} remains
true. This follows from the observation that we did not use any
locality constraints on the $n_i$'s or $\alpha(k_i,\pol_i)$ in
arriving at either \eqn{TwistIdentity2} or \eqn{invariance}. So one
can choose the $n_i$ to be nonlocal without affecting the value of
the amplitudes.  We can even set any one of the $n_i$ to zero.  For
example, choosing $\alpha(k_i,\pol_i) = n_u/u$ we get $n_u' = 0$.
This then puts a $u$ pole into the numerators $n'_s$ and $n'_t$,
making them nonlocal.  In fact this choice takes us back to the
relations found in the beginning of this section in
\eqn{FourPointRelation}. In the next section, we will obtain
nontrivial relations between higher-point tree-level partial
amplitudes, by choosing nonlocal numerators.

%%%%%%%%%%%%%%%%%%%%%%%%%%%%%%
\section{Higher-point generalization}
\label{NPointTwist}

In this section we generalize the relations found at four points to
higher points. To do this we will promote \eqn{TwistIdentity2} to be
the master identity of this paper.  We will use the close analogy
between this kinematic numerator identity and the color-group Jacobi
identity to apply it to higher-point tree-level amplitudes. We
will find that this is indeed possible given that certain extra
conditions are satisfied.  Specifically, given three dependent color
factors $c_\alpha, c_\beta, c_\gamma$ associated with tree-level color
diagrams, we propose that color-ordered scattering amplitudes can always be
decomposed into kinematic diagrams with numerator
factors $n_\alpha, n_\beta, n_\gamma$ that obey the analogous
numerator identity,
\begin{equation}
c_\alpha- c_\beta + c_\gamma = 0 \,,  \hskip 1cm \Rightarrow \hskip 1 cm   
n_\alpha - n_\beta + n_\gamma = 0\,.
\end{equation}

\subsection{The five-point kinematic identities}
\label{FivePointSubSection}

%%%%%%%%% FIGURE %%%%%%%%%%%%%%%%%%
\begin{figure}[t]
\centerline{\epsfxsize 5 truein \epsfbox{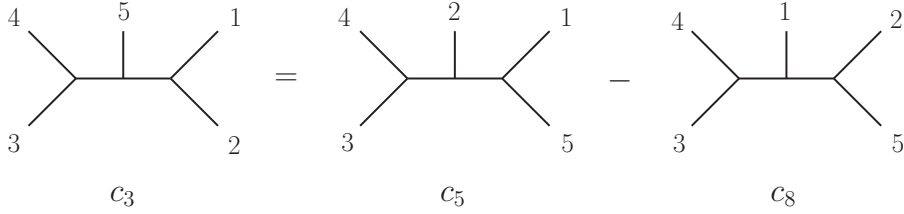}}
\caption[a]{\small The Jacobi identity at five points.  These diagrams
can be interpreted as relations for color factors, where each color
factor is obtained by dressing the diagrams with $\f^{abc}$ at each
vertex in a clockwise ordering.  Alternatively it can be interpreted
as relations between the kinematic numerator factors of corresponding
diagrams, where the diagrams are nontrivially rearranged compared to
Feynman diagrams.}
\label{FivePtTwistFigure}
\end{figure}
%%%%%%%%%%%%%%%%%%%%%%%%%%%%%%%%

First consider the five-point case.  We may again represent the
amplitudes in terms of diagrams with purely cubic interactions. These
diagrams specify the poles, examples of which are given in
\fig{FivePtTwistFigure}.  As in the four-point case we absorb any
contact terms into the numerator factors of these diagrams.  In total
there are 15 independent diagrams with two kinematic poles.  In order
for the four-point numerator identity to generalize we require that
the numerators of the 15 independent diagrams can be
arranged so that they satisfy exactly the same identities as the
$\f^{abc}$ composed color factors.

For example, consider the diagrams in
\fig{FivePtTwistFigure}. Interpreting these diagrams as color diagrams
we immediately see that as a consequence of the Jacobi identity they
satisfy the color-factor identity,
\begin{equation}
c_3 = c_5 - c_8 \, ,
\label{FivePointJacobi}
\end{equation}
where 
\begin{equation}
c_3  \equiv  \f^{a_3a_4b} \f^{ba_5c}\f^{ca_1a_2} \,, \hskip 1 cm 
c_5  \equiv  \f^{a_3a_4b} \f^{ba_2c}\f^{ca_1a_5} \,, \hskip 1 cm 
c_8  \equiv  \f^{a_3a_4b} \f^{ba_1c}\f^{ca_2a_5} \,.
\label{C358}
\end{equation}

If the kinematic numerators are to satisfy a corresponding identity
we must ensure that contact term contributions are associated with the
proper numerator.  This is not automatic, as we can easily check.
Even color-ordered Feynman diagrams at five points, in isolation, do
not generally satisfy the identity.  Rather the
numerator factors correspond to rearrangements
of the Feynman diagrams.

In general, a color-ordered five-point tree amplitude consist of five
diagrams,
\begin{equation}
A_5^\tree(1,2,3,4,5) = {n_1\over s_{12}s_{45}}+{n_2\over s_{23}s_{51}}+
{n_3\over s_{34}s_{12}}+{n_4\over s_{45}s_{23}}+{n_5\over s_{51}s_{34}} \,,
\end{equation}
where we use the notation $s_{ij}=(k_i+k_j)^2$, and the numerators are
simply labeled $n_i$ for $i=1,2,\ldots,15$.  Again we take the external 
momenta, $k_i$, to be massless and outgoing. The color-ordered
five-point amplitudes are symmetric under cyclic permutations and
antisymmetric under reflections. This means that there are at most 12
color orders that are not trivially related.  We have not yet
made use of the Kleiss-Kuijf relations~\cite{KleissKuijf,LanceColor}
given in \eqn{KleissKuijfFiveRelation}. Doing so reduces the number of
independent amplitudes down to only six,
\begin{eqnarray}
&&  A^\tree_5(1,2,3,4,5) \equiv {n_1\over s_{12}s_{45}}+{n_2\over s_{23}s_{51}}
+{n_3\over s_{34}s_{12}}+{n_4\over s_{45}s_{23}}+{n_5\over s_{51}s_{34}}\,, \nn\\
&&  A^\tree_5(1,4,3,2,5) 
         \equiv {n_{6}\over s_{14}s_{25}}+{n_5\over s_{43}s_{51}}
+{n_7\over s_{32}s_{14}}+{n_{8}\over s_{25}s_{43}}
+{n_2\over s_{51}s_{32}} \,, \nn\\
&&  A^\tree_5(1,3,4,2,5) \equiv {n_{9}\over s_{13}s_{25}}-{n_5\over s_{34}s_{51}}
+{n_{10}\over s_{42}s_{13}}-{n_{8}\over s_{25}s_{34}}
+{n_{11}\over s_{51}s_{42}}\,, \nn\\
&&  A^\tree_5(1,2,4,3,5) \equiv {n_{12}\over s_{12}s_{35}}
+{n_{11}\over s_{24}s_{51}}-{n_3\over s_{43}s_{12}}
+{n_{13}\over s_{35}s_{24}}-{n_5\over s_{51}s_{43}}, \nn\\
&&  A^\tree_5(1,4,2,3,5) \equiv {n_{14}\over s_{14}s_{35}}-
{n_{11}\over s_{42}s_{51}}-{n_7\over s_{23}s_{14}}-
{n_{13}\over s_{35}s_{42}}-{n_2\over s_{51}s_{23}} \,, \nn\\
&&  A^\tree_5(1,3,2,4,5) \equiv {n_{15}\over s_{13}s_{45}}-
{n_2\over s_{32}s_{51}}-{n_{10}\over s_{24}s_{13}}-{n_4\over s_{45}s_{32}}-
{n_{11}\over s_{51}s_{24}} \,,
\label{Indep5}
\end{eqnarray}
where the 15 numerator factors are distinguished by the propagator
structure that accompany them. The relative signs are again due to the
antisymmetry of the color-ordered Feynman vertices, or alternatively
the antisymmetry of the color factors.

It is important to note that the diagram expansion on
the right-hand side of \eqn{Indep5} will satisfy the Kleiss-Kuijf
relations. In fact, the Kleiss-Kuijf relations can be understood from
this perspective: Any color-ordered amplitudes that have an expansion
in terms of diagrams with only totally antisymmetric cubic vertices
will automatically satisfy the Kleiss-Kuijf relations.

The full color-dressed amplitudes can also be expressed 
in terms of the kinematic numerators $n_i$ and color factors via,
\begin{eqnarray}
{\cal A}^\tree_5& = & 
g^3\Bigl( {n_1 c_1\over s_{12}s_{45}}+{n_2 c_2\over s_{23}s_{51}}
+{n_3 c_3 \over s_{34}s_{12}}+{n_4 c_4\over s_{45}s_{23}}
+{n_5 c_5\over s_{51}s_{34}} +{n_{6} c_6\over s_{14}s_{25}}  \nn\\
&&\null \hskip .3 cm 
+ {n_7 c_7\over s_{32}s_{14}}+{n_{8} c_8\over s_{25}s_{43}}
+{n_{9} c_9\over s_{13}s_{25}}+{n_{10} c_{10}\over s_{42}s_{13}}
+{n_{11} c_{11}\over s_{51}s_{42}}+ {n_{12} c_{12}\over s_{12}s_{35}}  \nn\\
&&\null \hskip .3 cm 
+{n_{13} c_{13}\over s_{35}s_{24}} 
+{n_{14} c_{14}\over s_{14}s_{35}}
+{n_{15} c_{15}\over s_{13}s_{45}}\Bigr)\,,
\label{YangMillsNumerator}
\end{eqnarray}
where the color factors are explicitly given by,
\begin{eqnarray}
&& 
c_{1\phantom{0}} \equiv \f^{a_1 a_2 b}\f^{b a_3 c}\f^{c a_4 a_5}\,, \hskip 0.8cm  
c_{2\phantom{1}} \equiv \f^{a_2 a_3 b}\f^{b a_4 c}\f^{c a_5 a_1}\,, \hskip 0.8cm  
c_{3\phantom{1}} \equiv \f^{a_3 a_4 b}\f^{b a_5 c}\f^{c a_1 a_2}\,, \nn \\&&
c_{4\phantom{1}} \equiv \f^{a_4 a_5 b}\f^{b a_1 c}\f^{c a_2 a_3}\,, \hskip 0.8cm  
c_{5\phantom{1}} \equiv \f^{a_5 a_1 b}\f^{b a_2 c}\f^{c a_3 a_4}\,, \hskip 0.8cm  
c_{6\phantom{1}} \equiv \f^{a_1 a_4 b}\f^{b a_3 c}\f^{c a_2 a_5}\,, \nn \\&& 
c_{7\phantom{1}} \equiv \f^{a_3 a_2 b}\f^{b a_5 c}\f^{c a_1 a_4}\,, \hskip 0.8cm  
c_{8\phantom{1}} \equiv \f^{a_2 a_5 b}\f^{b a_1 c}\f^{c a_4 a_3}\,, \hskip 0.8cm  
c_{9\phantom{1}} \equiv \f^{a_1 a_3 b}\f^{b a_4 c}\f^{c a_2 a_5}\,, \nn \\&&
c_{10} \equiv \f^{a_4 a_2 b}\f^{b a_5 c}\f^{c a_1 a_3}\,, \hskip 0.8cm  
c_{11} \equiv \f^{a_5 a_1 b}\f^{b a_3 c}\f^{c a_4 a_2}\,, \hskip 0.8cm  
c_{12} \equiv \f^{a_1 a_2 b}\f^{b a_4 c}\f^{c a_3 a_5}\,, \nn \\\ &&
c_{13} \equiv \f^{a_3 a_5 b}\f^{b a_1 c}\f^{c a_2 a_4}\,, \hskip 0.8cm  
c_{14} \equiv \f^{a_1 a_4 b}\f^{b a_2 c}\f^{c a_3 a_5}\,, \hskip 0.8cm  
c_{15} \equiv \f^{a_1 a_3 b}\f^{b a_2 c}\f^{c a_4 a_5}\,. \hskip 1.5 cm 
\label{FivePointColor}
\end{eqnarray}
The number of independent color factors always coincide with the
number of partial amplitudes that are independent under the
Kleiss-Kuijf relations. This can, for example, be seen in the color
decomposition of ref.~\cite{LanceColor}. There are six independent
five-point partial amplitudes, corresponding to the six independent
color factors $c_i$.  The other color factors are related to these six
by the Jacobi identity.  At five points, this means that there can be
no more than six independent $n_i$'s, if the corresponding numerator
identity is to generalize. If these coincide---as they do---we might
expect that we should pick exactly one basis numerator per amplitude,
for example the numerator of the first term of every amplitude in
\eqn{Indep5}. However, this choice is likely not optimal for our
purposes. Naively, such a diagram basis would correspond to six choices of
gauge-dependent quantities which may not necessarily be mutually
compatible. If we instead pick $n_1,n_2, n_3, n_4$ and then define
$n_5$ and $n_6$, using,
\begin{eqnarray}
&& n_5 \equiv s_{51}s_{34}\Big( A_5^\tree(1,2,3,4,5)-
{n_1\over s_{12}s_{45}}-{n_2\over s_{23}s_{51}}-
{n_3\over s_{34}s_{12}}-{n_4\over s_{45}s_{23}}\Big) \,,
\label{n5TwistSolution}\\
&& n_6\equiv s_{14}s_{25}\Big( A_5^\tree(1,4,3,2,5)
 -{n_5\over s_{43}s_{51}}-{n_7\over s_{32}s_{14}}-{n_{8}\over s_{25}s_{43}}-
  {n_2\over s_{51}s_{32}}\Big) \,,
\label{n6TwistConstraint}
\end{eqnarray}
then fewer gauge-dependent choices are made, and the consistency of
amplitudes $A_5^\tree(1,2,3,4,5)$ and $A_5^\tree(1,4,3,2,5)$ is
automatically guaranteed. (Here we have defined $n_6$ in terms of
$n_7$ and $n_8$ but this will not cause any problems, as they turn out
to depend only on $n_1,n_2,n_3,n_4$.)

We require that the remaining numerators, $n_7, n_8, \ldots, n_{15}$,
satisfy the same Jacobi identity equations as the corresponding color
factors associated with each diagram.  For example, we require,
\begin{equation}
c_3 - c_5 + c_8 = 0\,,  \hskip 1cm \Rightarrow \hskip 1cm   
n_3 - n_5 + n_8=0\,,
\label{NumeratorColorEquivalence}
\end{equation}
where the color factors and numerators correspond to 
\eqn{YangMillsNumerator},  and the explicit values of the color
factors are given in \eqn{FivePointColor}.
Working our way through all possible Jacobi identities at
five points we find ten numerator identities,
\begin{eqnarray}
&& n_3 - n_5 + n_8=0 \,,  \nn\\
&& n_3 - n_1 + n_{12}=0\,,  \nn\\
&& n_4 - n_1 + n_{15}=0\,,  \nn\\
&& n_4 - n_2 + n_7=0\,,  \nn\\
&& n_5 - n_2 + n_{11}=0\,,  \nn\\
&& n_7 - n_6 + n_{14}=0\,,  \nn\\
&& n_8 - n_6 + n_9=0\,,  \nn\\
&& n_{10} - n_9 + n_{15}=0\,,  \nn\\
&& n_{10} - n_{11} + n_{13}=0\,, \nn\\
&&  \Bigl(n_{13} -n_{12} + n_{14} = 0\Bigr)\,,
\label{twisteqns}
\end{eqnarray}
where the last equation is redundant.
Solving for the nine numerators, $n_7, n_8, \ldots, n_{15}$, then gives,
\begin{eqnarray}
&&  n_7 = n_2 - n_4\,,  \nn\\
&&  n_8 = -n_3 + n_5\,,  \nn\\
&&  n_9 = n_3 - n_5 + n_6\,,  \nn\\
&&  n_{10} = -n_1 + n_3 + n_4 - n_5 + n_6\,,  \nn\\
&&  n_{11} = n_2 - n_5\,,  \nn\\
&&  n_{12} = n_1 - n_3\,,  \nn\\
&&  n_{13} = n_1 + n_2 - n_3 - n_4 - n_6\,,  \nn\\
&&  n_{14} = -n_2 + n_4 + n_6\,, \nn\\
&&  n_{15} = n_1 - n_4\,.
\label{TwistSolution}
\end{eqnarray}

We may now replace  $n_7$ and $n_8$ in our definition of $n_6$, 
\begin{eqnarray}
n_6 &=& A_5^\tree(1, 4, 3, 2, 5) s_{14} s_{25}
- A_5^\tree(1, 2, 3, 4, 5) (s_{15} + s_{25}) s_{14}
+ n_1  {s_{14} (s_{15} + s_{25}) \over s_{12} s_{45}} \nn\\
&& \null 
+ n_2 {s_{23} + s_{35} \over s_{23}}
+ n_3  {s_{14} \over s_{12}}
+ n_4  {s_{14} s_{15} + s_{14} s_{25} + s_{25} s_{45} \over s_{23}  s_{45}} \,.
\label{n6TwistSolution}
\end{eqnarray}
Note that both $n_5$ and $n_6$ appear to be nonlocal quantities in
general, but by adjusting $n_1,n_2,n_3,n_4$ in \eqn{n6TwistSolution}
we can cancel poles in $A_5(1,2,3,4,5)$ and $A_5(1,4,3,2,5)$, making
all numerators local.

Any one of the five-point color-ordered tree amplitudes $A_5^\tree$
can now be found by writing down the diagram expansion and plugging in
the solutions of the corresponding numerator factors from the
numerator identities (\ref{n5TwistSolution}), (\ref{TwistSolution})
and (\ref{n6TwistSolution}).  The amplitudes will be functions of six
parameters: the four gauge-dependent numerators $n_1,n_2, n_3, n_4$
and the two gauge-independent amplitudes $A_5^\tree(1,2,3,4,5),
A_5^\tree(1,4,3,2,5)$.  However, since $n_1,n_2,n_3, n_4$ are gauge
dependent this construction may seem problematic. Indeed we must check
that the construction is consistent with the known properties of
amplitudes.

We verify consistency by considering all factorization channels of the
five-point tree amplitudes. In these limits each five-point amplitude
factorizes into the product of a three-point and a four-point
amplitude.  In the factorization channels the five-point numerator
equations~(\ref{twisteqns}) reduce to the four-point numerator
identity ~(\ref{TwistIdentity2}). Thus, any potential violation of the
general numerator identity must come from contact terms that vanish in
all of these limits. Similarly, as the five-point numerator equations
are gauge invariant in these factorization channels, the relevant
terms that potentially break the gauge invariance of the five-point
identity must be exactly these contact terms. Therefore, we need to
investigate how the five-point numerators behave under the freedom,
\begin{eqnarray}
&& n_1'=n_1+\alpha_1 s_{12}s_{45} \,, \hskip 1cm
n_2'=n_2+\alpha_2 s_{23}s_{51} \,, \nn \\
&& n_3'=n_3+\alpha_3 s_{34}s_{12}\,, \hskip 1cm
 n_4'=n_4+\alpha_4 s_{45}s_{23} \,.
\end{eqnarray}
Let us, for the moment, treat the $\alpha$'s as local objects, but
otherwise arbitrary functions of the kinematics and polarizations. By
construction $A_5^\tree(1,2,3,4,5)$ and $A_5^\tree(1,4,3,2,5)$ are
invariant under this freedom, since $n_5$ and $n_6$ are constrained so
as to correctly reproduce these amplitudes. From the definition of $n_5$
and $n_6$ we obtain the transformations,
\begin{eqnarray}
n_5' =&&  n_5 - (\alpha_1 + \alpha_2 + \alpha_3 + \alpha_4) 
             s_{51} s_{34} \,, \nn \\
n_6' =&& n_6 + \alpha_3 s_{12} s_{14} +  
 (\alpha_1 + \alpha_2 + \alpha_3 + \alpha_4) s_{14}s_{15} \nn\\
&& \null
+ (\alpha_1 + \alpha_3 + \alpha_4) s_{14} s_{25} - 
        \alpha_2 s_{15}s_{25} + \alpha_4 s_{25} s_{45} \,.
\end{eqnarray}

Remarkably, we find the four remaining amplitudes in \eqn{Indep5}
are also invariant under these transformations. Plugging in the shifts
of numerators $n_1, n_2, \ldots, n_6$ into the remaining $n_i$ using
\eqn{TwistSolution} gives a set of nontrivial cancellations,
\begin{eqnarray}
&&  \Delta A_5^\tree(1,3,4,2,5)={\Delta n_{9}\over s_{13}s_{25}}-
{\Delta n_5\over s_{34}s_{51}}+{\Delta n_{10}\over s_{42}s_{13}}-
{\Delta n_{8}\over s_{25}s_{34}}+{\Delta n_{11}\over s_{51}s_{42}}=0 \,, \nn\\
&& \Delta  A_5^\tree(1,2,4,3,5)={\Delta n_{12}\over s_{12}s_{35}}+
{\Delta n_{11}\over s_{24}s_{51}}-{\Delta n_3\over s_{43}s_{12}}+
{\Delta n_{13}\over s_{35}s_{24}}-{\Delta n_5\over s_{51}s_{43}}=0 \,, \nn\\
&&  \Delta A_5^\tree(1,4,2,3,5)={\Delta n_{14}\over s_{14}s_{35}}-
{\Delta n_{11}\over s_{42}s_{51}}-{\Delta n_7\over s_{23}s_{14}}-
{\Delta n_{13}\over s_{35}s_{42}}-{\Delta n_2\over s_{51}s_{23}}=0 \,, \nn\\
&& \Delta A_5^\tree(1,3,2,4,5)={\Delta n_{15}\over s_{13}s_{45}}-
{\Delta n_2\over s_{32}s_{51}}-{\Delta n_{10}\over s_{24}s_{13}}-
{\Delta n_4\over s_{45}s_{32}}-{\Delta n_{11}\over s_{51}s_{24}}=0 \,,
\hskip .5 cm 
\end{eqnarray}
where $\Delta n_i = n_i'-n_i$.   

As for the four-point case, there is no need to restrict the $\alpha$
parameters to be local. In fact we can pick the $\alpha$'s so that
$n'_1=n'_2=n'_3=n'_4=0$, which implies that the explicit dependence on
these parameters in the amplitudes cancel out. Hence, our construction
is completely gauge invariant since the amplitudes in \eqn{Indep5}
depend only on the two gauge-invariant basis amplitudes $A_5^\tree(1,
2, 3, 4, 5)$ and $A_5^\tree(1, 4, 3, 2, 5)$.

Feeding the numerator solutions in
eqs.(\ref{n5TwistSolution}), (\ref{TwistSolution}) and
(\ref{n6TwistSolution}) into \eqn{Indep5}, we find remarkably simple
relations,
\begin{eqnarray}
&& A_5^\tree(1, 3, 4, 2, 5)={-s_{12} s_{45} A_5^\tree(1, 2, 3, 4, 5)+
 s_{14} (s_{24} + s_{25}) A_5^\tree(1, 4, 3, 2, 5) \over s_{13}s_{24}}\,,\nn \\
&& A_5^\tree(1, 2, 4, 3, 5)={-s_{14} s_{25} A_5^\tree(1, 4, 3, 2, 5) + 
  s_{45} (s_{12} + s_{24}) A_5^\tree(1, 2, 3, 4, 5) \over s_{24}s_{35}}\,,\nn \\
&& A_5^\tree(1, 4, 2, 3, 5)= {-s_{12} s_{45} A_5^\tree(1, 2, 3, 4, 5) + 
  s_{25} (s_{14} + s_{24}) A_5^\tree(1, 4, 3, 2, 5) \over s_{35}s_{24}}\,,\nn \\
&& A_5^\tree(1, 3, 2, 4, 5)= {-s_{14} s_{25} A_5^\tree(1, 4, 3, 2, 5) + 
    s_{12} (s_{24} + s_{45}) A_5^\tree(1, 2, 3, 4, 5) \over s_{13}s_{24}} \,, 
\hskip 1 cm 
 \label{FivePointRelations}
\end{eqnarray}
independent of $n_1,n_2,n_3,n_4$.  Thus, we find novel nontrivial
relations between color-ordered gauge-theory tree amplitudes.  Note
that these relations should hold for {\it any} helicity configuration
and they should be valid in $D$ dimensions. We
explicitly verified these for $D$-dimensional five-gluon
amplitudes. However, we would also like to have a general argument as
to why these hold.

This is found by looking at the factorization limits more
carefully. The five-point numerator equations reduce to the correct
gauge-invariant four-point identity in all factorization
limits. Furthermore we know that at four points we can make the
numerators local (for example, by using individual color-ordered
Feynman diagrams).  Similarly, we can make the five-point numerators
local, by construction.  Specifically, from \eqn{n5TwistSolution}, we
can make $n_5$ local by taking $n_1,n_2,n_3,n_4$ to be the
coefficients of the corresponding poles in the amplitude.  When these
poles are subtracted, the only remaining poles can be
$1/s_{51}s_{34}$, so that $n_5$ is local. It is a bit trickier to show
that $n_6$ is local. The $n_5$ and $n_2$ terms are automatically the
correct terms for subtracting poles in $A_5(1,4,3,2,5)$.  However,
$n_7$ and $n_8$ cannot be adjusted since we demand that these satisfy
\eqn{TwistSolution}.  Do they have the correct values to subtract the
poles? Indeed, they do.  This is because in any factorization limit
the identities in \eqn{TwistSolution} do hold, since we already
demonstrated that they hold at four points. Thus $n_6$ is also local.
The only missing pieces of the amplitudes that cannot be seen in
factorization limits are the five-point contact terms. These
potentially missing pieces are, of course, local.  However from
dimensional analysis, a five-point gauge-theory amplitude cannot contain a
five-point contact term.  Such a term would correspond to a forbidden
five-point contact term in the Lagrangian.  Thus we can always find a
local solution to numerator factors satisfying the identity.  The
resulting amplitudes have the correct factorizations limits, strongly
suggesting the consistency of our construction.

We expect this argument to generalize also to higher-point tree amplitudes,
allowing us to do similar rearrangements of the terms in the
amplitudes such that the respective kinematic numerator identities are
satisfied, with local or nonlocal numerators. What remains to be
determined is how much freedom we have in this rearrangement, that is,
how many of the numerator factors can be left undetermined and thus
gauge dependent, and how many numerators are constrained by extra
constraints of the type \eqn{n6TwistConstraint}. In the next section
we will identify the all-$n$ pattern.

\subsection{Implications for $n$ points}
\label{NPointSection}

As with the lower-point cases, we expect the numerator
identity~(\ref{TwistIdentity2}) to lead to new constraints between
amplitudes for any number of external legs. Under the Kleiss-Kuijf
relations we know that there are at most $(n-2)!$ independent
color-ordered amplitudes.  Here we argue that the kinematic numerator
identity imposes
additional constraints so that the number of independent color-ordered
amplitudes is $(n-3)!$.

%%%%%%%%%%%% TABLE %%%%%%%%%%%%%%%%%%%%%%%%%%
\def\hs{\hskip .2 cm \null }
\begin{table*}
\caption{Counts of various diagrams, equations and amplitudes as a
function of the number of external points. We count only diagrams with
three vertices.  The first row gives the number of diagrams that
appear in color-ordered partial amplitudes (where the external legs
are cyclically ordered).  The second row gives the number of such
diagrams that appear in a full color-dressed amplitude. The third row
gives the number of numerator (or equivalently Jacobi identity)
equations.  The fourth row gives the number of such independent
equations.  The fifth row gives the number of linearly independent (or
basis) numerators which are not constrained by these equations. The
number of independent amplitudes under the Kleiss-Kuijf relations is
given in the sixth row. The number of independent basis partial
amplitudes under the new kinematic identity, in terms of which all
others can be expressed, is given in the last row. The last row is a
conjecture beyond eight points.}
\label{NPointTable} 
\vskip .4 cm
\begin{tabular}{|l|*{6}{c}|c|}
\hline
external legs  	&  $\phantom{i}$ 3 $\phantom{i}$ &  $\phantom{i}$ 4 $\phantom{i}$ &  $\phantom{x}$ 5$\phantom{x}$&  $\phantom{x}$ 6$\phantom{x}$& $\phantom{xx}$  7  $\phantom{xx}$ &  $\phantom{xx}$ 8 $\phantom{xx}$ &  $n$ \\
\hline
ordered diagrams		& 1 & 2 & 5 & 14 & 42 & 132 &
                                         $\frac{2^{n-2}(2n-5)!!}{(n-1)!}$\\
diagrams 	& 1 & 3 & 15 & 105 & 945 & $10\;395$ & 
                                                           $(2n-5)!!$\\ 
numerator equations   & 0 & 1 & 10 & 105 & 1260 &  $17\, 325$ &
                                         $\frac{(n-3)(2n-5)!!}{3}$\\
indep. numerator eqs.      & 0 & 1 & 9 & 81& 825&  $9\,675$  & 
                                \hs\hs  {\scriptsize $(2n-5)!!-(n-2)!$}\hs\\
basis numerators  & 1 & 2 & 6 & 24 &  120 &  720 & $(n-2)!$  \\
Kleiss-Kuijf amplitudes  \hs & 1 & 2 & 6 & 24 &  120 &  720  &  
                                                            $(n-2)!$  \\
basis amplitudes	& 1 & 1 & 2 & 6 &  24 &  120  & $(n-3)!$\\
\hline
\end{tabular}
\vskip .5 cm
\end{table*}

%%%%%%%%%%%%%%%%%%%

Following the four- and five-point discussion we may expand each of
these color-ordered amplitudes in terms of diagrams with only cubic
and totally antisymmetric vertices, that is, numerators and
propagators, 
\begin{equation}
A^\tree_n(1,2,3,\ldots,n)=\sum_{j} {n_j\over (\prod_{m} p^2_m)_j}\,,
\end{equation}
where the sum runs over all distinct ordered diagrams.
The number of color-ordered diagrams with a fixed ordering of external
legs given in the first row in table~\ref{NPointTable}.  For this we
count diagrams with only cubic vertices, we absorb
four-point contact terms into numerator factors that cancel
propagators.  The total number of distinct diagrams at $n$-points
contributing to the full color-dressed amplitudes is $(2n-5)!!$ as
listed in the second row of the table.

Applying the same reasoning as to lower points we have explored
higher-point properties of the kinematic identity. This leads to the
following conjecture for the $n$-point structure:
\begin{enumerate}

\item The kinematic numerators of gauge-theory tree-level diagrams can
always be rearranged to satisfy the numerator
identity~(\ref{TwistIdentity2}) equations, which is in one-to-one
correspondence to the Jacobi identity equations satisfied by the color
factors of the same diagrams. The numerators can be either local or
nonlocal.

\item One must simultaneously rearrange the diagrams of at least
  $(n-3)!$ partial amplitudes to ensure gauge invariance of the full
  amplitude.  The remaining partial amplitudes will
  automatically satisfy the numerator equations, since they are built
  from the same diagrams.

\end{enumerate}

In Table~\ref{NPointTable}, we have collected various numerical counts
helpful for understanding the effect of the numerator identity at higher
points. The description of the count in each row is given in the
caption.  For all numbers given in the table we have explicitly
constructed the count.  The $n$-point count in the last row remains
a conjecture, beyond eight points.

The above conjecture does not address how this arrangement of diagrams
is best achieved. 
The approach we take here is to simply treat the numerators as
being unknown variables satisfying an equation system that describes
the conjecture,
\begin{eqnarray}
 \{n_\alpha&=&n_\beta-n_\gamma\}
   \label{TwistIdentity3}\, ,
\\&&\nn\\
A^\tree_n(\Perm_i\{1,2,3,\ldots,n\})&=&
\left[\sum_{j} {n_j\over (\prod_{m} p^2_m)_j}\right]_i\label{constraint}\, ,
   \end{eqnarray}
where \eqn{TwistIdentity3} represents all possible numerator
identities that can be written down at the $n$-point level
corresponding to the color-factor Jacobi identities, and
\eqn{constraint} represents the statement that the diagrams dressed
with the unknown numerators must sum up to the known partial
amplitudes for at least $(n-3)!$ different permutations of
the external legs labeled by $i=1,\ldots,(n-3)!$.

The solution to these equations will give kinematic numerators that
are functions of a set of $(n-3)!$ basis amplitudes. The basis
amplitudes must be chosen to be independent under the Kleiss-Kuijf
relations, since these relations are manifest in our diagrammatic
representation, but otherwise the basis is arbitrary.  As noted in
table~\ref{NPointTable}, there are a total $(2n-5)!!$ diagrams and
associated numerators.  By counting the equations we see that the
solution of \eqn{TwistIdentity3} and \eqn{constraint} will not fix all
these numerators, but it will leave $(n-2)!-(n-3)!$
unspecified. 
However, our conjecture implies that any amplitude 
built out of the solution to \eqn{TwistIdentity3} and
\eqn{constraint} will be independent of these free numerators, and
will only depend on the gauge-invariant basis amplitudes.  In 
particular, we can set the unspecified numerators to zero without 
altering any partial amplitudes.  We have explicitly checked
this through eight points.

Specifically, we solve \eqn{TwistIdentity3} and \eqn{constraint} using
basis amplitudes where legs $1,2,3$ are fixed
$A^\tree_n(1,2,3,\Perm\{4,\ldots,n\})$. By feeding the solved
numerators into those amplitudes not in the basis, we obtain
new tree-level relations. As part of this conjecture we expect that 
these relations hold for the partial amplitudes with any external
polarizations in $D$ dimensions. 

Using the fact that the Kleiss-Kuijf relations allows us to always put
legs 1 and 2 next to each-other, we need only give the formula for the
case where leg 3 is separated from leg 2 by a set of legs $\{\alpha\}$
and similarly separated from leg 1 by a set 
$\{\beta\}$.
We order the leg labels in $\{\alpha\}$ and $\{\beta\}$ as,
\begin{equation}
\{\alpha\} \equiv \{4,5,\ldots,m-1,m\}, \hskip 1 cm \{\beta\} \equiv 
\{m+1,m+2,\ldots,n-1,n\} \,,
\end{equation}
which can always be undone in the final expressions by a
permutation of legs $4,\ldots,n$.  By extrapolating from the structure
of the solutions evaluated up through eight external particles, we obtain
an all-$n$ form,
\begin{equation}
A_n^\tree(1,2,\{\alpha\},3,\{\beta\})=
\sum_{\{\sigma\}_j \in {\rm POP}(\{\alpha\},\{\beta\})} 
A_n^\tree(1,2,3,\{\sigma\}_j)  \prod_{k=4}^m  
{{\cal F}(3,\{\sigma\}_j,1| k)\over s_{2,4,\ldots,k} } \,,
\label{alln}
\end{equation}
where the sum runs over ``partially ordered permutations'' (POP) of
the merged $\{\alpha\}$ and $\{\beta\}$ sets. This corresponds to all
permutations of $\{\alpha\} \bigcup \{\beta\}$ that maintains the
order of the $\{\beta\}$ elements. Either set may be taken as empty,
but if $\{\alpha\}$ is empty the equation becomes trivial. The 
function ${\cal F}$ associated with leg $k$ is given by,
\begin{eqnarray}
{\cal F}(3,\sigma_1,\sigma_2, \ldots, \sigma_{n-3},1| k) 
     \equiv {\cal F} (\{\rho\}| k)&=&
\left\{ 
\begin{array}{ll}
          \sum_{l=t_k}^{n-1} {\cal G}(k,\rho_l) & \mbox{if $t_{k-1} < t_{k}$}\\
        - \sum_{l=1}^{t_k} {\cal G}(k,\rho_l) & \mbox{if $t_{k-1} > t_{k}$}
\end{array} \right\} \nn \\&&
 \null +
\left\{ 
\begin{array}{ll}
         s_{2,4,\ldots,k} & \mbox{if $t_{k-1} < t_{k}< t_{k+1}$}\\
        -s_{2,4,\ldots,k} & \mbox{if $t_{k-1} > t_{k} > t_{k+1}$}\\
         0 & \mbox{else}
\end{array} \right\} \,,
\hskip 1 cm 
\end{eqnarray}
and where $t_k$ is the position of leg $k$ in the set $\{\rho\}$,
except for $t_3$ and $t_{m+1}$ which are always defined to
be\footnote{An alternative choice is  $t_3\equiv \infty, t_{m+1} \equiv
0$ which is equivalent to \eqn{BoundaryConditions} by momentum conservation.},
\begin{equation}
t_3\equiv t_5 \,, \hskip 2 cm t_{m+1} \equiv 0 \,.
\label{BoundaryConditions}
\end{equation}
(Note that for $m=4$ this implies $t_3=t_{m+1}=0$). 
The function ${\cal G}$ is given by,
\begin{equation}
{\cal G }(i,j)=\left\{ 
\begin{array}{ll}
        s_{i,j}  & \mbox{if $i< j$ or $j=1,3$}\\
        0& \mbox{else}
\end{array} \right\} \,.
\end{equation}
Finally, the kinematic invariants are,
\begin{equation}
s_{i,j}=(k_i+k_j)^2, \hskip 1 cm s_{2,4,\ldots,i}=(k_2+k_4+\ldots+k_i)^2,
\end{equation}
where the momenta are massless and outgoing.  We have explicitly
confirmed in $D=4$ that all MHV amplitudes through 12 points satisfy
\eqn{alln}.  We have also checked that gluon amplitudes for all helicity 
configurations through eight points satisfy this.

The four-, five- and six-point relations generated by this formula are, 
\begin{eqnarray} 
A_4^\tree(1,2, \{4\}, 3) &=& {A_4^\tree(1, 2, 3, 4) s_{14} \over s_{24}} \,,
\nn\\ \nn\\
A_5^\tree(1,2, \{4\}, 3,\{5\}) &=& 
{ A_5^\tree(1,2, 3, 4,5) (s_{14} + s_{45})+
  A_5^\tree(1, 2, 3, 5, 4) s_{14} \over s_{24}}\,,\phantom{xxxxxxxx}
\nn\\ \nn\\
A_5^\tree(1, 2, \{4,5\}, 3) &=& {-A_5^\tree(1, 2, 3, 4, 5) s_{34} s_{15} 
  - A_5^\tree(1, 2, 3, 5, 4) s_{14} (s_{245}+s_{35}) \over s_{24} s_{245}}\,,
       \nn\\\nn\\
A_6^\tree(1, 2, \{4\}, 3, \{5, 6\}) &=& 
{A_6^\tree(1, 2, 3, 4, 5, 6) (s_{14} + s_{46}+s_{45})\over s_{24}} \nn\\ 
&& \null 
+{A_6^\tree(1, 2, 3, 5, 4, 6) (s_{14} + s_{46}) \over s_{24}}
+{A_6^\tree(1, 2, 3, 5, 6, 4) s_{14} \over s_{24}}\,, \nn\\\nn\\
A_6^\tree(1,2,\{4,5\},3,\{6\}) &=& 
- {A_6^\tree(1,2,3,4,5,6)s_{34} (s_{15} + s_{56})\over s_{24} s_{245}} \nn\\
&& \null 
- {A_6^\tree(1,2,3,4,6,5) s_{34}s_{15} \over s_{24} s_{245}}\nn\\
&& \null
- {A_6^\tree(1,2,3,6,4,5) (s_{34} + s_{46})s_{15} \over s_{24} s_{245}} \nn\\
&& \null 
-{A_6^\tree(1,2,3,5,4,6) (s_{14} + s_{46})(s_{245}+ s_{35})\over s_{24} 
      s_{245}}\nn\\
&& \null
-{A_6^\tree(1,2,3,5,6,4) s_{14} (s_{245}+ s_{35})\over s_{24} s_{245}} \nn \\
&& \null
-{A_6^\tree(1,2,3,6,5,4) s_{14} 
   (s_{245}+ s_{35}+s_{56})\over s_{24} s_{245}} \,, \nn\\ \nn\\
A_6^\tree(1,2,\{4,5,6\},3)&=& 
-{A_6^\tree(1,2,3,4,5,6) s_{34} (s_{245}+s_{56}+s_{15})s_{16} 
\over s_{24} s_{245}s_{2456}}  \nn\\
&& \null 
+ {A_6^\tree(1,2,3,4,6,5) s_{34}s_{15}  
(s_{2456} + s_{36})\over s_{24} s_{245}s_{2456}}  \nn\\
&& \null 
+ {A_6^\tree(1,2,3,6,4,5) (s_{34} + s_{46}) s_{15} 
(s_{2456} + s_{36})\over s_{24} s_{245}s_{2456}} \nn\\
&& \null 
- {A_6^\tree(1,2,3,5,4,6) (s_{14} + s_{46})s_{35} s_{16}\over s_{24} 
s_{245}s_{2456}} \nn\\
&& \null 
-{A_6^\tree(1,2,3,5,6,4) s_{14} s_{35}s_{16}  
   \over s_{24} s_{245}s_{2456}} \nn\\
&& \null
+ {A_6^\tree(1,2,3,6,5,4) s_{14} (s_{245}+s_{35}+s_{56} )
                (s_{2456} + s_{36}) \over s_{24} s_{245}s_{2456}} \,.
\label{6point}
\end{eqnarray}
We introduce the brackets ``$\{$'' and ``$\}$'' 
to emphasize  the connection to \eqn{alln}---they carry no other significance.

One amusing point is that this solution allows us to express any
partial amplitude with three negative helicities as a linear
combination of the ``split helicity'' cases where the three negative 
helicity legs are nearest neighbors in the color
ordering~\cite{SplitHelicity}.

As for four- and five-point amplitudes, we expect the higher-point numerator
identities to lead to interesting consequences at loop level, via the
unitarity method. In the next section we will address these
consequences, albeit only for the simplest case of the four-point
identity.

%%%%%%%%%%%%%%%%%%%%%%%%
\section{Higher-loop applications}
\label{HigherLoopSection}

%%%%%%%%% FIGURE %%%%%%%%%%%%%%%%%%
\begin{figure}[t]
\centerline{\epsfxsize 5.5 truein \epsfbox{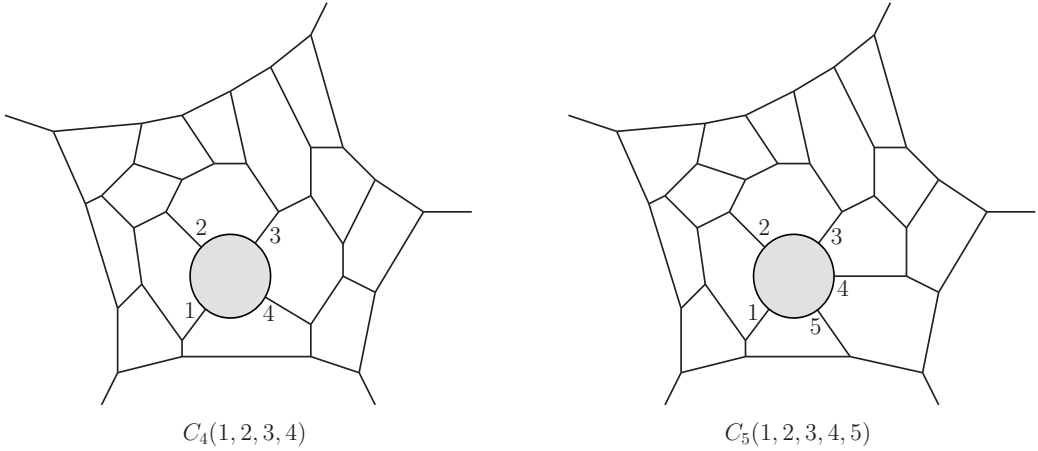}}
\caption[a]{\small Near-maximal cuts with specified color order given
by the uncut blob. All visible lines are cut, thus on-shell. The blobs
are tree amplitudes with implied sums over all helicity and particle
types entering and leaving the blobs.}
\label{MaximalCutTwistFigure}
\end{figure}
%%%%%%%%%%%%%%%%%%%%%%%%%%%%%%%%

The four-point identity (\ref{TwistIdentity2}) appears to be rather
innocuous compared to the higher-point relations, but as we now
discuss even this case has interesting consequences at higher loops.

\subsection{Near-maximal cuts}
Consider a near-maximal cut where we leave only one four-point tree
blob uncut, but where all other other propagators are cut, as
illustrated in \fig{MaximalCutTwistFigure}(a). As in \sect{FourPoint},
we are interested in relating different color orderings, hence we
label the cut similarly,
\begin{equation}
C_4(1,2,3,4) \sim A_4^\tree(1,2,3,4) \,.
\end{equation}
To be specific we can work with the cut in
\fig{MaximalCutTwistFigure}(a), but the structure of the cut outside
of the four-point blob is unimportant, since it will play no role in
the analysis.  The blob appearing in the cut, up to the polarizations,
spins and particle types being summed over, is equivalent to a
color-ordered four-point tree amplitude. Therefore, if the tree
amplitudes satisfy the relations given in the previous section, the
cuts must as well,
\begin{eqnarray}
&& \hskip 2.2 cm  C_4(1,2,3,4) + C_4(1,3,4,2) + C_4(1,4,2,3) =0 \,, \nn \\
&& t\, C_4(1,2,3,4)=u\, C_4(1,3,4,2) \,,  
\hskip 1cm s\, C_4(1,2,3,4) = u\, C_4(1,4,2,3)\,, \nn \\ &&
\hskip 3.5 cm t\, C_4(1,4,2,3)=s \, C_4(1,3,4,2)\,,
\label{cutidentities}
\end{eqnarray}
where the Mandelstam variables are now understood to be for the cut
internal loop momenta, $s=(l_1+l_2)^2$, $t=(l_2+l_3)^2$ and
$u=(l_1+l_3)^2$, not the external momenta.

We can use the factorization properties of the four-point tree 
amplitudes appearing in the blob to express the cut as,
\begin{equation}
C_4(1,2,3,4)= \frac{n_s}{s}+\frac{n_t}{t}\,,  \hskip 1. cm 
C_4(1,3,4,2)= -\frac{n_u}{u}-\frac{n_s}{s} \,, \hskip 1. cm 
C_4(1,4,2,3)= \frac{n_u}{u}-\frac{n_t}{t} \,,
\label{STUcut}
\end{equation}
where the numerators $n_s, n_t$ are the $s$-channel and $t$-channel
numerators, respectively.  Following the same steps as for the
tree-level discussion in \sect{FourPoint} we arrive at the same
kinematic identity for the numerators,
\begin{equation}
n_u = n_s - n_t \,.
\label{TwistIdentity2Cut}
\end{equation}
As for the tree-level case, the numerator factors in the cuts also have
a freedom similar to \eqns{SgaugeInvariance}{TUgaugeInvariance}. 

%%%%%%%%% FIGURE %%%%%%%%%%%%%%%%%%
\begin{figure}[t]
\centerline{\epsfxsize 5.5 truein \epsfbox{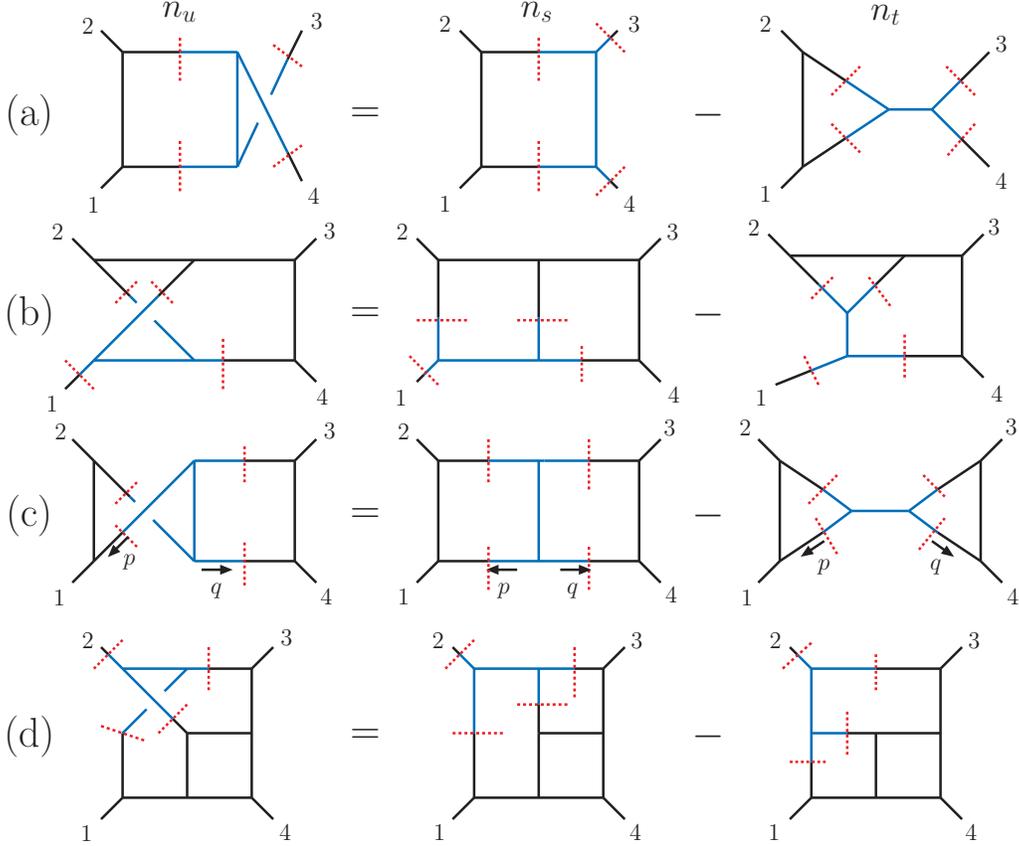}}
\caption[a]{\small Applications of the four-point numerator identity
to loop amplitudes.  The equalities should be interpreted either as
equalities for the color factors obtained by dressing the diagrams
with $\f^{abc}$s at each vertex, or as equalities between numerators
of cut diagrams and hold for any gauge theory. All lines are cut
except for the propagator inside the (blue) four-point diagram
indicated by the dotted lines.  Our construction here is similar to
\fig{MaximalCutTwistFigure}, where we consider near-maximal cuts.  }
\label{diagTwistExamplesFigure}
\end{figure}
%%%%%%%%%%%%%%%%%%%%%%%%%%%%%%%%

We can now apply the four-point identity
(\ref{TwistIdentity2Cut}) to any  higher-loop amplitude.  In
\fig{diagTwistExamplesFigure} we give various examples of such
applications. The idea is that if one computes the numerator 
contributions of the diagrams on the right hand side, all numerator
terms are determined on the left hand side, up to the cut conditions.
The diagrams in the figure specify the propagator
structure of contributions under study.  We expect the relations
in \fig{diagTwistExamplesFigure} to hold for any gauge theory. 
In each of these diagrams the associated color factors are just 
those obtained by dressing each diagram by $\f^{abc}$'s at each vertex.

We note that the construction presented here based on near-maximal
cuts with a four-point blob, as shown in
\fig{MaximalCutTwistFigure}(a), generalizes to higher-point blobs. For
example, we can consider instead a near-maximal cut with a five-point
blob, as illustrated in \fig{MaximalCutTwistFigure}(b). This cut obeys
the same identities and relations as the the five-point amplitude
$A_5^\tree(1,2,3,4,5)$ in \sect{FivePointSubSection}.  We can continue
to relax the cut conditions obtaining further generalizations based on
higher-point blobs.  With further relaxation of cut conditions we can
also have generalized cuts with several independent tree blobs. Each
$n$-point blob will now have its own $n$-point identity and
relations. In general, any generalized cut will contain nontrivial
relations between the constituent diagrams analogous to the tree-level
relations presented in previous sections.  

As simple checks of the relations, we used the one-~\cite{GSB},
two-\cite{BRY,BDDPR} and three-loop~\cite{BRY,GravityThree} expressions
for the four-point amplitudes of $\NeqFour$ super-Yang-Mills theory to
confirm all examples in \fig{diagTwistExamplesFigure}.  To see how
this works in practice with more physical theories, here we work
out a two-loop example in QCD.

\subsection{Two-loop QCD examples}

%%%%%%%%% FIGURE %%%%%%%%%%%%%%%%%%
\begin{figure}[t]
\centerline{\epsfxsize 5 truein \epsfbox{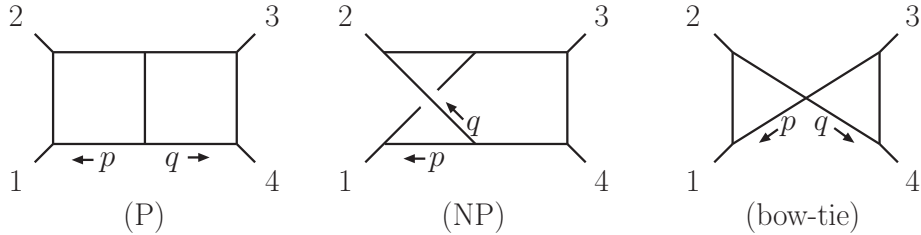}}
\caption[a]{\small The diagrams corresponding to integrals contributing
to the identical-helicity two-loop four-point QCD amplitude. The numerator
factors are given in the text.}
\label{TwoLoopIntegralsFigure}
\end{figure}
%%%%%%%%%%%%%%%%%%%%%%%%%%%%%%%%

As a nontrivial example of the four-point numerator identity in QCD
we consider the two-loop four-gluon all-plus helicity amplitude of
pure Yang-Mills theory.  This amplitude has been worked out in
ref.~\cite{TwoLoopAllPlus} in terms of the color decomposition given
in \eqn{FColor}.  The result is that the planar primitive amplitude
is,
\begin{eqnarray}
A^\P_{1234}
& =& i \, { \spb1.2\spb3.4 \over \spa1.2\spa3.4 }
 \biggl\{
 s_{12} \, I_4^\P \Bigl[ (D_s-2) ( \mud_p^2 \, \mud_q^2 
         + \mud_p^2 \, \mud_{p+q}^2  + \mud_q^2 \, \mud_{p+q}^2 )  
       \nn\\
&& \hskip 4 cm \null 
       + 16 \Bigl( (\mud_p \cdot \mud_q)^2 - \mud_p^2 \, \mud_q^2 \Bigr) 
                    \Bigr](s_{12},s_{23}) \nn \\
&& \null \hskip 2.6 cm 
+ 4 (D_s-2) \, I_4^\bowtie[(\mud_p^2 + \mud_q^2) 
                \, (\mud_p \cdot \mud_q) ] (s_{12}) \nn \\
&& \null \hskip 2.6 cm
+ {(D_s-2)^2 \over s_{12}} \, I_4^\bowtie\Bigl[ 
      \mud_p^2 \, \mud_q^2 \, ( (p+q)^2 + s_{12} ) \Bigr] (s_{12},s_{23})
              \biggr\} \,.
 \label{ppppPlanar}
\end{eqnarray}
Similarly, the nonplanar primitive amplitude is, 
\begin{equation}
A^\NP_{12;34}
= i \, { \spb1.2\spb3.4 \over \spa1.2\spa3.4 } \, s_{12} 
I_4^\NP \Bigl[ (D_s-2) ( \mud_p^2 \, \mud_q^2 
                   + \mud_p^2 \, \mud_{p+q}^2 
                   + \mud_q^2 \, \mud_{p+q}^2 )
+ 16 \Bigl( (\mud_p \cdot \mud_q)^2
           - \mud_p^2 \, \mud_q^2 \Bigr) \Bigr](s_{12},s_{23})\,.
\label{ppppNonPlanar}
\end{equation}
The vectors $\vec\mud_p$, $\vec\mud_q$ represent the
$(-2\eps)$-dimensional components of the loop momenta $p$ and $q$
and $(D_s-2)$ counts the number of gluon states circulating in 
the loop---in four dimensions $D_s = 4$.  
The planar double-box integral, whose corresponding diagram is
depicted in \fig{TwoLoopIntegralsFigure}(a) is defined as
\begin{eqnarray}
&& I_4^\P [N (\mud_i, p,q,k_i)] (s_{12},s_{23}) 
 \equiv \int
 {d^{D}p\over (2\pi)^{D}} \,
 {d^{D}q\over (2\pi)^{D}}\, \nn \\
&& \hskip 2 cm \null \times
  { {N} (\mud_i, p,q,k_i) \over 
     p^2\, q^2\, (p+q)^2 (p - k_1)^2 \,(p - k_1 - k_2)^2 \,
        (q - k_4)^2 \, (q - k_3 - k_4)^2 }  \,, \hskip 1 cm 
\label{PlanarInt} 
\end{eqnarray}
where ${N} (\mud_i, p,q,k_i)$ represents the numerator factor, which
is a polynomial in the momenta.  Similarly, the nonplanar double-box
integral, depicted in \fig{TwoLoopIntegralsFigure}(b), is given by
\begin{eqnarray}
&& 
I_4^\NP [{N} (\mud_i, p,q,k_i)] (s_{12},s_{23}) 
 \equiv \int \! {d^{D} p \over (2\pi)^{D}} \,
        {d^{D} q \over (2\pi)^{D}}\, \nn \\
&& \hskip 2.5 cm \null \times 
{{N} (\mud_i, p, q, k_i) \over p^2\, q^2\, (p+q)^2 \,
         (p-k_1)^2 \,(q-k_2)^2\,
   (p+q+k_3)^2 \, (p+q+k_3+k_4)^2}\,. \hskip .9 cm 
\label{NonPlanarInt}
\end{eqnarray}
Note that $A^\NP_{12;34}$ is symmetric under $k_1 \leftrightarrow k_2$, and 
under $k_3 \leftrightarrow k_4$.   
The ``bow-tie'' integral $I_4^\bowtie$ shown in
\fig{TwoLoopIntegralsFigure}(c) is defined by
\begin{eqnarray}
&& I_4^{\bowtie} [{N}  (\mud_i, p,q,k_i)] (s_{12})
 \equiv \int {d^{D}p\over (2\pi)^{D}} \,
 {d^{D}q\over (2\pi)^{D}}\, \nn\\
&& \hskip 2 cm \null \times 
 { {N} (\mud_i, p,q,k_i) \over 
     p^2\, q^2\, (p - k_1)^2 \,(p - k_1 - k_2)^2 \,
        (q - k_4)^2 \, (q - k_3 - k_4)^2 }\,.
\label{BowTieInt} 
\end{eqnarray}
Note that in the color decomposition~(\ref{FColor}), the color factors
associated with the bow-tie integral are those of the parent 
double-box integral.

The amplitude has a simple feature which is rather
surprising from the Feynman diagram point of view: the planar 
and nonplanar numerators have the same structure.  We may use
the cut-loop numerator identity displayed in
\fig{diagTwistExamplesFigure}(b), to explains this feature.
To compare the numerators we need to relabel the cut
momenta in the nonplanar double-box integral so that it matches the
cut momenta of the planar double-box integral.  This relabeling
amounts to swapping $\mud_q \leftrightarrow \mud_{p+q}$ in the
nonplanar integrals.  Since the nonplanar numerator is invariant
under this swap, we may directly read off the numerator from the
nonplanar contribution (\ref{ppppNonPlanar}).  Up to an overall
prefactor the numerator is,
\begin{equation}
n_u = (D_s-2) s_{12} ( \mud_p^2 \, \mud_q^2 + \mud_p^2 \, \mud_{p+q}^2 
                   + \mud_q^2 \, \mud_{p+q}^2 )
+ 16 s_{12} \Bigl( (\mud_p \cdot \mud_q)^2 - \mud_p^2 \, \mud_q^2 \Bigr) \,. 
\end{equation}
Similarly from the planar integral we may read off the numerator
factor for the planar integral.  This then gives us the numerator,
\begin{equation}
 n_s = (D_s-2) s_{12} (\mud_p^2 \, \mud_q^2
+ \mud_p^2 \, \mud_{p+q}^2 + \mud_q^2 \, \mud_{p+q}^2 ) + 16 s_{12}\Bigl(
(\mud_p \cdot \mud_q)^2 - \mud_p^2 \, \mud_q^2 \Bigr) \,.
\end{equation}
The bow-tie integrals do not contribute to the indicated cuts in
\fig{diagTwistExamplesFigure}(b).  Since there is no contribution with
propagators corresponding to the second diagram on the right-hand-side
of \fig{diagTwistExamplesFigure}(c) the corresponding numerator
vanishes,
\begin{equation}
 n_t  = 0\,.
\end{equation}
Thus we see that the numerator identity (\ref{TwistIdentity2Cut})
corresponding to \fig{diagTwistExamplesFigure}(b) is indeed satisfied.
It also explains the previously mysterious identical structures of the
nonplanar and planar double-box numerators.

We may also use the numerator identity, as applied in
\fig{diagTwistExamplesFigure}(c) to constrain the bow-tie
contributions in $A^\P_{1234}$.  First consider the terms proportional to 
$(D_s - 2)$. Since there are no $1/s_{12}$ contributions we may take 
\begin{equation}
n_t = 0\,,
\label{Nteqn}
\end{equation}
where $n_t$ corresponds to the last term in
\fig{diagTwistExamplesFigure}(c) (where $s,t,u$ refer to the figure and
not the external kinematics).  Reading off $n_s$ from $A_{1234}$ we have,
\begin{eqnarray}
n_{s} &=& s_{12} \, (D_s-2) (\mud_p^2 \, \mud_q^2 
         + \mud_p^2 \, \mud_{p+q}^2  + \mud_q^2 \, \mud_{p+q}^2 ) 
 + 16 s_{12}\Bigl( (\mud_p \cdot \mud_q)^2 - \mud_p^2 \, \mud_q^2 \Bigr)\nn \\
&& \null \hskip 2 cm 
       +  4 (D_s-2) \, (\mud_p^2 + \mud_q^2) 
                \, (\mud_p \cdot \mud_q)  (p+q)^2 \,, 
\end{eqnarray}
where the last term comes from the $D_s -2$ bow-tie contribution.
Similarly from $A^\P_{2134} = A^\P_{3421}$ we can read off,
\begin{eqnarray}
n_{u} &=&  s_{12} \, (D_s-2) ( \mud_p^2 \, \mud_q^2 
         + \mud_p^2 \, \mud_{p-q}^2  + \mud_q^2 \, \mud_{p-q}^2 ) 
 + 16 s_{12}\Bigl( (\mud_p \cdot \mud_q)^2 - \mud_p^2 \, \mud_q^2 \Bigr) \nn \\
&& \hskip 2 cm \null
       -  4 (D_s-2) \, (\mud_p^2 + \mud_q^2) 
                \, (\mud_p \cdot \mud_q)  (p-q+k_3+k_4)^2 \,,
\end{eqnarray}
where we adjusted the momentum labels to ensure that the cut momenta are
the same as in $A_{1234}$.  We have
\begin{eqnarray}
n_s - n_u &=& 4 (D_s-2) \, (\mud_p^2 + \mud_q^2) 
                \, (\mud_p \cdot \mud_q) \Bigl( s_{12} 
                  +  (p+q)^2 + (p-q+k_3+k_4)^2 \Bigr)\nn \\
  & =& 0 \,, 
\end{eqnarray}
where we made use of the cut conditions.  Thus, from here and from
\eqn{Nteqn}, we find that the numerator identity
(\ref{TwistIdentity2Cut}) is satisfied.

We have also confirmed that the $(D_s -2)^2$ terms of the bow-tie
contributions satisfy the numerator identity in
\fig{diagTwistExamplesFigure}(c).  This is somewhat trickier than for
the other terms, because the color factors of the terms with a
$1/s_{12}$ pole need to be rearranged so they correspond to the last
diagram of \fig{diagTwistExamplesFigure}(c), not the planar double-box
diagrams, before the identity is apparent.

%%%%%%%%%%%%%%%%%%%%%%%%%%%%%%%%%%%%%%%%%%%%%%

The systematics of how we use the numerator
identity~(\ref{TwistIdentity2}) in conjunction with the method of
maximal cuts~\cite{FiveLoop} will be discussed
elsewhere~\cite{FourLoopNonplanar}.  For the special case of
$\NeqFour$ super-Yang-Mills theory, various additional relations
between different contributions, including some between different loop
orders, may be found in refs.~\cite{BRY,BDDPR,FiveLoop,Freddy}.

%%%%%%%%%%%%%%%%%%%%%%%%%%%

\section{Implications for gravity amplitudes}
\label{GravitySection}

The KLT relations~\cite{KLT} tell us that gravity tree amplitudes can
be expressed directly in terms of gauge-theory tree amplitudes.  These
relations were originally derived in string theory and hold in field
theory, since the low-energy limit of string theory is field theory.
However, from a purely field-theoretic viewpoint, starting from the
Einstein-Hilbert and Yang-Mills Lagrangians, these relations have
remained obscure~\cite{BernGrant}.  Some new relations between gravity
and gauge-theory MHV amplitudes were recently presented in
ref.~\cite{Freedman1}, adding to the mystery.

In this section we use the identity (\ref{TwistIdentity2}) to
clarify the relationship between gravity and gauge theories, arguing
that the KLT relations are equivalent to a diagram-by-diagram
numerator ``squaring'' relation with gauge theory.

Consider first the four-point color-dressed amplitude in
\eqn{FullTree}. As already noted in ref.~\cite{BDDPR}, the four-point
gravity tree amplitudes can be expressed directly in terms of diagrams
whose numerators are squares of the numerators of the corresponding gauge
theory.  In particular, for pure gravity we have,
\begin{equation}
-iM_4^\tree(1,2,3,4)= \frac{n_s^2}{s}+
                           \frac{n_t^2}{t}+\frac{n_u^2}{u}\,,
\end{equation}
where the numerators $n_i$ are just the kinematic numerators of the gauge
theory.  More generally, for other particle contents, the two gauge-theory 
amplitudes corresponding to each factor of $n_i$ can be
different.  Distinguishing the two gauge-theory 
amplitudes with a tilde, we have
\begin{equation}
-iM^\tree_4(1,2,3,4)= \frac{n_s \n_s}{s}+
                        \frac{n_t \n_t}{t}+\frac{n_u \n_u}{u}\,.
\end{equation}
It is straightforward to verify that this expression reproduces the
correct amplitude using the four-point KLT form in \eqn{KLT4} together
with the definition of the numerators in \eqn{PoleForm}.  It is essential
here to use the fact that the gauge-theory numerators satisfies the
identities,
\begin{equation}
n_u = n_s -n_t \,, \hskip 2 cm 
\n_u = \n_s - \n_t \,.
\label{TwistEqsGrav}
\end{equation}

Can we generalize this behavior to higher points?  Indeed it is
straightforward to check that the five-point KLT relation
(\ref{KLT5}) is equivalent to a sum over all fifteen diagrams defined
in \eqn{Indep5}, but with a product of two gauge-theory numerators,
\begin{eqnarray}
-iM^\tree_5(1,2,3,4,5) & = & 
{n_1 \n_1\over s_{12}s_{45}}+{n_2 \n_2\over s_{23}s_{51}}
+{n_3 \n_3 \over s_{34}s_{12}}+{n_4 \n_4\over s_{45}s_{23}}
+{n_5 \n_5\over s_{51}s_{34}} +{n_{6} \n_6\over s_{14}s_{25}}  \nn\\
&&\null
+ {n_7 \n_7\over s_{32}s_{14}}+{n_{8} \n_8\over s_{25}s_{43}}
+{n_{9} \n_9\over s_{13}s_{25}}+{n_{10} \n_{10}\over s_{42}s_{13}}
+{n_{11} \n_{11}\over s_{51}s_{42}}+ {n_{12} \n_{12}\over s_{12}s_{35}}  \nn\\
&&\null
+{n_{13} \n_{13}\over s_{35}s_{24}} 
+{n_{14} \n_{14}\over s_{14}s_{35}}
+{n_{15} \n_{15}\over s_{13}s_{45}} \,.
\label{GravityNumerator}
\end{eqnarray}
Again, for this to hold, it is important that the $n_i$'s 
satisfy the numerator identities in \eqn{twisteqns}, and that
the $\n_i$'s satisfy corresponding ones.

Using \eqn{GravityNumerator} we can obtain new relations between
gravity and gauge-theory amplitudes simply by altering the basis
amplitudes when solving for the kinematic numerators $n_i, \n_i$.  For
example, if we use $A^\tree_5(1, 2, 3, 4, 5)$ and $A^\tree_5(1,3,2, 4,
5)$ for the basis amplitudes for the left (tilde-less) gauge theory
and $\widetilde A^\tree_5(2, 1, 4, 3, 5)$ and $\widetilde A^\tree_5(3,
1, 4, 2, 5)$ for the basis amplitudes for the right (tilde) gauge
theory, we immediately obtain the KLT relation (\ref{KLT5}) from
\eqn{GravityNumerator}.  On the other hand if we change the basis
amplitudes of both the left and right gauge theories to same orderings,
$(1, 2, 3, 4, 5)$ and $(1,4,3,2,5)$, we find an alternative left-right
symmetric representation of the five-point gravity amplitudes,
\begin{eqnarray}
-i M_5^\tree(1,2,3,4,5) 
  &=& {s_{12} s_{45} (s_{12} s_{14} s_{23} + s_{34}
 (s_{12} + s_{13}) (s_{23} + s_{25})) \over s_{13} s_{24} s_{35}} \nn  \\
&& \hskip 2 cm \null
\times  A_5^\tree(1,2,3,4,5) \widetilde A_5^\tree(1,2,3,4,5)\nn\\
&& \null 
  -{s_{12} s_{14} s_{25} (s_{13} + s_{35}) s_{45} \over s_{13} s_{24} s_{35}}
   \Bigl(A_5^\tree(1,2,3,4,5) \widetilde A_5^\tree(1,4,3,2,5)\nn\\
&& \null \hskip 4.5 cm 
 +   A_5^\tree(1,4,3,2,5)  \widetilde{A}_5^\tree(1,2,3,4,5) \Bigr)\nn\\
&&\null 
 + {s_{14} s_{25} (s_{12} s_{14} s_{34} 
  +s_{23} (s_{13} + s_{14}) (s_{34} + s_{45}))
          \over s_{13} s_{24} s_{35}} \nn\\
&& \hskip 1 cm \null \times 
 A_5^\tree(1,4,3,2,5)
               \widetilde A_5^\tree(1,4,3,2,5) \,.
\label{NewFivePointKLT}
\end{eqnarray}
This representation can also be obtained directly from the KLT relations by
substituting in the expressions for the gauge-theory amplitudes
in terms of the same basis partial amplitudes used in
\eqn{NewFivePointKLT}.

In general, we expect the numerator relation between gravity
and gauge theories to hold for an arbitrary number of external
legs. Starting from the color-dressed gauge-theory amplitudes,
\begin{eqnarray}
&& {1\over g^{n-2}}{\cal A}^\tree_n(1,2,3,\ldots,n)=\sum_{i}
                   {n_i c_i \over (\prod_{j} p^2_j)_i}\, , \nn \\
&&  {1\over g^{n-2}} {\cal \widetilde{A}}^\tree_n(1,2,3,\ldots,n)=\sum_{i}
                   {\n_i c_i \over (\prod_{j} p^2_j)_i} \, ,
\label{GaugeTheoryLeftRight}
\end{eqnarray}
where the $n_i$ and $\n_i$ satisfy the kinematic numerator
identity~(\ref{TwistIdentity2}), the $c_i$ satisfy the Jacobi
identity, and where the sum runs over all diagrams with only cubic
vertices.  We then expect gravity amplitudes to be given by,
\begin{eqnarray}
&&  -iM^\tree_n(1,2,3,\ldots,n)=\sum_{i} 
                   {n_i \n_i \over (\prod_{j} p^2_j)_i} \, ,
\label{GravityMasterFormula}
\end{eqnarray}
where the sum runs over the same set of diagrams as in
\eqn{GaugeTheoryLeftRight}.  We have explicitly confirmed that this is
consistent with the KLT relations through eight points.  We may think
of formula (\ref{GravityMasterFormula}) as a master formula for
generating new representations of gravity amplitudes. As part of the
conjecture, this equation has a freedom corresponding to the $2 (n-3)
(n-3)!$ numerators $n_i$ and $\n_i$ not fixed by any constraints and
which drop out of the amplitudes~(\ref{GaugeTheoryLeftRight})
and~(\ref{GravityMasterFormula}).  Furthermore, if we solve for
the $n_i$, $\n_i$ in terms of gauge-theory basis amplitudes, as done
in section~\ref{NPointSection}, then one has the freedom to choose
these $2(n-3)!$ basis amplitudes. Every choice would result in a
different KLT-like relation when fed into the master
formula~(\ref{GravityMasterFormula}).  The five-point
representation~(\ref{NewFivePointKLT}) is one example of many possible
KLT-like relations between gravity and gauge-theory amplitudes that we
can construct. Note that the number of tilde and no-tilde gauge-theory
amplitudes appearing in the $n$-point KLT relations (\ref{KLTnpoint})
is $(n-3)!$ each, matching the number of independent gauge-theory
amplitudes. Indeed, imposing the choice of basis amplitudes
corresponding to the ones appearing in the KLT relations,
\begin{equation}
A_n^\tree(1,\Perm\{2,\ldots,n-2\},n-1,n)\,, \hskip 1. cm 
{\tilde A}_n^\tree(\Perm\{i_1,\ldots,i_j\},1,n-1,
                                \Perm\{l_1,\ldots,l_{j'}\},n)\,,
\end{equation}
where $\Perm$ signifies all permutations over the arguments,
recovers the original KLT relations.

We expect that the simplified connection between gravity and gauge
theory tree amplitudes presented here should make it easier to link
the ultraviolet properties of gravity theories to those of gauge
theories.

%%%%%%%%%%%%%%%%%%%%%%%%%%%%%%%%%%%%%%%%%%%%%%%%%%

\section{Conclusions}

In this paper we presented a new kinematic identity for $n$-point
tree-level color-ordered gauge-theory amplitudes.  This identity is
the kinematic analog of the Jacobi identity for color. By solving the
constraints imposed by the identity we obtained nontrivial relations
between tree-level color-ordered partial amplitudes. We derived the
relevant identities at four and at five points in some detail and have
confirmed it explicitly for all gluon helicity amplitudes through
eight points.  Beyond this it remains a conjecture, although we have
performed a variety of consistency checks.  A consequence of this
identity is that it gives nontrivial relations between different
color-ordered tree amplitudes.  We conjectured an explicit all-$n$
formula relating the different color-ordered $n$-point amplitudes.
Under the Kleiss-Kuijf relations between color-ordered partial
amplitudes~\cite{KleissKuijf}, for a given helicity and particle
configuration there are $(n-2)!$ independent partial amplitudes.  The
new relations reduce this number to $(n-3)!$ independent partial
amplitudes.

Using generalized unitarity we demonstrated that this kinematic
numerator identity implies nontrivial relations between contributions
at higher loops.  In this paper we applied these relations to the QCD
two-loop four-point amplitude with identical helicities to explain a
previously mysterious similarity between planar and nonplanar
contributions.  This amplitude is much simpler than for the other
helicity configurations in QCD, but it does serve to illustrate the
constraints imposed on higher-loop gauge-theory amplitudes by the
numerator identity.

We also discussed the implication of the kinematic numerator identity for
gravity tree amplitudes.  We found that though at least eight points
this numerator identity, together with the KLT relations, imply that
gravity tree-level amplitudes can be put in a diagrammatic form where
each numerator is a product of two corresponding gauge-theory
numerators.  We conjecture that this can be done for any number of
external particles.  Using the solution of color-ordered gauge-theory
amplitudes in terms of a basis set of such amplitudes, we also showed
how to rearrange the KLT relations into new forms.

A natural arena for applying these identities is in high-loop studies
of maximally supersymmetric gauge and gravity amplitudes.  As will be
discussed elsewhere~\cite{FourLoopNonplanar}, these are very useful
for obtaining four-loop four-point amplitudes in $\NeqFour$
super-Yang-Mills theory, including the subleading color contributions.
The super-Yang-Mills amplitudes are useful both for studying AdS/CFT
conjecture~\cite{ABDK} and as input to the corresponding calculations
in $\NeqEight$ supergravity to determine their ultraviolet
properties~\cite{BDDPR, Finite, GravityThree}, which appear better
behaved than anticipated, and may even be finite.  Indeed,
cancellations appear to continue to all loop orders in a class of
terms detectable in certain cuts~\cite{Finite}. These cancellation
follow from the existence of novel one-loop
cancellations~\cite{NoTriangle}.  These cancellations do not appear to
be connected to supersymmetry.  Instead, they appear connected to the
recently uncovered behavior~\cite{GravityTreeRecursion} of gravity
tree amplitudes under large complex
deformations~\cite{NoTriCancellations}.  String dualities have also
been used to argue for improved ultraviolet behavior of $\NeqEight$
supergravity, though various difficulties with decoupling towers of
massive states may spoil this conclusion~\cite{DualityArguments}.

It would be very interesting to explore the consequences of the
kinematic numerator identity for spontaneously broken and massive theories,
especially in theories of phenomenological interest.  More generally,
we expect the new relations discussed in this paper to be helpful for
further clarifying the structure of perturbative gauge and gravity
theories.  In particular, we expect it to aid higher-loop
investigations of gauge and gravity theories using the unitarity
method.

\section*{Acknowledgements}
\vskip -.3 cm 

We thank Lance J. Dixon, Harald Ita, David A. Kosower and Radu Roiban
for many helpful discussions and collaboration on related topics.  We
also thank Fernando Febres Cordero and Darren Forde for helpful
discussions.  We thank Academic Technology Services at UCLA for
computer support.  This research was supported by the US Department of
Energy under contracts DE--FG03--91ER40662 and DE--AC02--76SF00515.
J.~J.~M.~C.  and H.~J. gratefully acknowledge the financial support
of Guy Weyl Physics and Astronomy Alumni Fellowships.

%%%%%%%%%%%%%%%%%%%%%%%%%%%%%%%%%%%%%%%%%%%%%%%%%%%%%%%%%%

\end{document}